\documentclass[9pt,twocolumn,twoside]{pnas-new}
\usepackage{xr}

\graphicspath{{./MainFigures/}}
\usepackage{siunitx}
\usepackage{overpic}
\usepackage{upgreek}

\templatetype{pnasresearcharticle} % Choose template
\begin{document}

\title{Quantum relaxometry for detecting biomolecular interactions with single NV centers}

% Use letters for affiliations, numbers to show equal authorship (if applicable) and to indicate the corresponding author
\author[a,b,c]{Min Li}
\author[c,d,1]{Qi Zhang}
\author[e]{Xi Kong}
\author[a,b]{Sheng Zhao}
\author[f]{Bin-Bin Pan}
\author[a,b]{Ziting Sun}
\author[a,b]{Pei Yu}
\author[c]{Zhecheng Wang}
\author[a,b]{Mengqi Wang}
\author[a,b]{Wentao Ji}
\author[a,b,g]{Fei Kong}
\author[a,b,g]{Guanglei Cheng}
\author[h]{Si Wu}
\author[a,b,g]{Ya Wang}
\author[a,b,c]{Sanyou Chen}
\author[f,1]{Xun-Cheng Su}
\author[a,b,c,g,1]{Fazhan Shi}

\affil[a]{School of Physical Sciences, University of Science and Technology of China, Hefei 230026, China}
\affil[b]{Anhui Province Key Laboratory of Scientific Instrument Development and Application, University of Science and Technology of China, Hefei 230026, China}
\affil[c]{School of Biomedical Engineering and Suzhou Institute for Advanced Research, University of Science and Technology of China, Suzhou 215123, China}
\affil[d]{Institute of Quantum Sensing, School of Physics, Institute of Fundamental and Transdisciplinary Research, Zhejiang Key Laboratory of R$\&$D and Application of Cutting-edge Scientifc Instruments, State Key Laboratory of Ocean Sensing, Zhejiang University, Hangzhou 310027, China}
\affil[e]{National Laboratory of Solid State Microstructures and Department of Physics, Nanjing University, Nanjing 210093, China}
\affil[f]{State Key Laboratory of Elemento-organic Chemistry and College of Chemistry, Nankai University, Tianjin 300071, China}
\affil[g]{Hefei National Laboratory, University of Science and Technology of China, Hefei 230088, China}
\affil[h]{Hefei National Research Center for Physical Sciences at the Microscale, Department of Polymer Science and Engineering, University of Science and Technology of China, Hefei 230026, China}

% Please give the surname of the lead author for the running footer
\leadauthor{Shi}

% Please add a significance statement to explain the relevance of your work
\significancestatement{Quantum sensing using nitrogen-vacancy (NV) centers in diamond promisingly offers spin observation window to detect biomolecular interactions at single-molecule level. A critical bottleneck, however, lies in achieving high-density molecular immobilization on the diamond surface without compromising bioactivity---a challenge stemming from inefficient surface functionalization and steric interference. We overcome this limitation by engineering a nanogel-coated interface that enables dense molecular anchoring while preserving native physiological functions. This innovation not only enables observation of single-molecule interactions but also resolves ensemble-averaged interaction behaviors at the microscale. By bridging the gap between single-molecule stochasticity and population-level biophysics, our platform establishes a versatile framework for high-precision analysis of molecular interactions, with broad implications for proteomic studies, diagnostics, and the fundamental understanding of biological processes.}

% Please include corresponding author, author contribution and author declaration information
\authorcontributions{F.S. supervised the project; Q.Z., and F.S. designed research; M.L., Q.Z., X.K., S.Z., B.P., Z.S., P.Y., Z.W., M.W., W.J., F.K., G.C., S.W., Y.W., S.C., X.S., and F.S. performed experiment; M.L., Q.Z., X.K., and F.S. analyzed data; and Q.Z., M.L., and F.S. wrote the paper.}
\authordeclaration{The authors declare no competing interest.}
%\equalauthors{\textsuperscript{1}A.O.(Author One) contributed equally to this work with A.T. (Author Two) (remove if not applicable).}
\correspondingauthor{\textsuperscript{1}To whom correspondence should be addressed. \\
E-mail: \textcolor{blue}{zhq2011@ustc.edu.cn}; \textcolor{blue}{xunchengsu@nankai.edu.cn}; \textcolor{blue}{fzshi@ustc.edu.cn}\\}

% At least three keywords are required at submission. Please provide three to five keywords, separated by the pipe symbol.

\renewcommand{\sectionautorefname}{\textit{SI Appendix} Sec.}%
\renewcommand{\subsectionautorefname}{\textit{SI Appendix} Sec.}%

\begin{abstract}
The investigation of biomolecular interactions at the single-molecule level has emerged as a pivotal research area in life science, particularly through optical, mechanical, and electrochemical approaches. Spins existing widely in biological systems, offer a unique degree of freedom for detecting such interactions. However, most previous studies have been largely confined to ensemble-level detection in the spin degree. Here, we developed a molecular interaction analysis method approaching single-molecule level based on relaxometry using the quantum sensor, nitrogen-vacancy (NV) center in diamond. Experiments utilized an optimized diamond surface functionalized with a polyethylenimine nanogel layer, achieving $\sim$10 nm average protein distance and mitigating interfacial steric hindrance. Then we measured the strong interaction between streptavidin and spin-labeled biotin complexes, as well as the weak interaction between bovine serum albumin and biotin complexes, at both the micrometer scale and nanoscale. For the micrometer-scale measurements using ensemble NV centers, we re-examined the often-neglected fast relaxation component and proposed a relaxation rate evaluation method, substantially enhancing the measurement sensitivity. Furthermore, we achieved nanoscale detection approaching single-molecule level using single NV centers. This methodology holds promise for applications in molecular screening, identification and kinetic studies at the single-molecule level, offering critical insights into molecular function and activity mechanisms.
\end{abstract}

%\dates{This manuscript was compiled on \today}
%\doi{\url{www.pnas.org/cgi/doi/10.1073/pnas.XXXXXXXXXX}}

\maketitle
\thispagestyle{firststyle}
\ifthenelse{\boolean{shortarticle}}{\ifthenelse{\boolean{singlecolumn}}{\abscontentformatted}{\abscontent}}{}

\firstpage[8]{2}
% Use \firstpage to indicate which paragraph and line will start the second page and subsequent formatting. In this example, there are a total of 11 paragraphs on the first page, counting the first level heading as a paragraph. The value {12} represents the number of the paragraph starting the second page. If a paragraph runs over onto the second page, include a bracket with the paragraph line number starting the second page, followed by the paragraph number in curly brackets, e.g. "\firstpage[4]{11}".

% If your first paragraph (i.e. with the \dropcap) contains a list environment (quote, quotation, theorem, definition, enumerate, itemize...), the line after the list may have some extra indentation. If this is the case, add \parshape=0 to the end of the list environment.
\dropcap{S}tudying molecular interactions and dynamics at the single-molecule level is of vital significance for exploring the intrinsic mechanisms of biological and chemical processes. To date, a variety of single-molecule measurement techniques have been developed, utilizing approaches like optics, mechanics and electrochemistry. Notable examples include single-molecule fluorescence resonance energy transfer (smFRET)\cite{fret2008natmeth,fret2022natcomm}, fluorescence correlation spectroscopy (FCS)\cite{FCS2021,FCS2023Ana}, Raman spectroscopy\cite{Raman2017single,raman2024single}, high-speed atomic force microscopy (AFM)\cite{hsAFM2011,hsAFM2019,hsAFM2023}, optical and magnetic tweezers\cite{opticaltweezers2014,magnetweezers2008,forcespec2015}, and nanostructured devices based on nanopores\cite{nanopore2019proteinsingle,nanopore2022single,nanopore2024DNAsingle} and nanowires\cite{CNTsinglemol2022PNAS,NT2018single,CNT2016singleDNA}. These techniques have been extensively applied to single-molecule measurements, including studies of DNA hybridization and melting kinetics\cite{FRET2018,FET2017ncSINGLEDNA}, protein binding\cite{plasmonic2020singleprotein}, and enzyme activity\cite{CNTsinglemol2022PNAS}. However, optical-based methods suffer from photobleaching, limiting the long-term observation. Electrochemical methods based on nanopores and nanowires are susceptible to interference from ionic concentrations in the solution\cite{debyescreenCNT2011}. Furthermore, an emerging trend in single-molecule research is the integration of multiple complementary physical measurement windows to acquire more comprehensive and precise insights into molecular interactions\cite{tweezerOpt2019nanotech,ElectroOptical2022smallstruc}.

Spin-based measurement methods are promising in overcoming the challenges faced by the aforementioned optical and electrochemical approaches. Biological systems inherently possess diverse spin sources, such as nuclear and electron spins in proteins, free radicals involved in metabolic activities, and transition metal ions, which collectively provide a broad signal bandwidth\cite{zhangt2021biosens-acssens}. While traditional magnetic resonance (MR) techniques excel at ensemble-level analysis, they lack the sensitivity for single-molecule detection. The nitrogen-vacancy (NV) center has garnered extensive attention in recent years as a quantum sensor, due to its excellent magnetic field sensitivity and outstanding biocompatibility. Notably, individual biomolecule detections based on spin signals have been achieved using NV centers\cite{nanolett2014Gdsingle,SFZ2015science,nuclear2016science,DNA2018kfnm}. This advancement provides both technical infrastructure and conceptual frameworks for detecting molecular interactions at this single-molecule level. NV-based quantum relaxometry, with simple pulse sequences, has been successfully applied to detect magnetic noise of free radicals\cite{AcsNano2020ChemReact,acsSENSOR2020H2O2,Jacs2022H2O2,sciadv2021radical,redox2022radical,acscent2023radical,acsnano2023radical,Nanolett2023radical,funcdiamond2024radical,PNAS2024radical}, pH\cite{pH2017chemevent,pH2019ACSnano}, and metal-based paramagnetic spins such as Gd$^{3+}$, Mn$^{2+}$ and Cu$^{2+}$ ions \cite{nanolett2014Gdsingle,2013pnasGdlipid,nanolett2013Mn,natcomm2017Cu,SciAdv2024cellforce,CommChem2024mRNA,wpf2019sciadFe,nanolett2024Fe,PRA2024zc}. These metal-based spin labels offer photostability and sub-nanometer dimensions superior to fluorescent labels, presenting great potential in molecular interaction detection. 

Up to now, there is still a lack of effective attempts to measure biomolecular interactions at the single-molecule level in the spin degree. In this work, we presented a biomolecular interaction analysis method that integrates metal-based spin labeling with quantum relaxometry employing NV centers in diamond. The basis of this methodology is the precise engineering of the sensor-biomolecule interface, which optimizes bonding density while preserving molecular activity. By functionalizing the diamond surface with a polyethylenimine (PEI) nanogel, we achieved a protein bonding distance of $\sim$10 nm and minimized steric hindrance. A novel relaxation rate evaluation method, reconsidering the often-overlooked fast relaxation component, was developed to improve the measurement sensitivity of ensemble NV centers. This enabled sensitive detection of both strong and weak interactions between biomolecules at the micrometer scale. We further extended this approach to single NV centers, achieving measurements at nanoscale and near single-molecule level.

\section*{Results}
\begin{figure*}[hbtp]
\centering
\begin{overpic}[width=.5\textwidth]{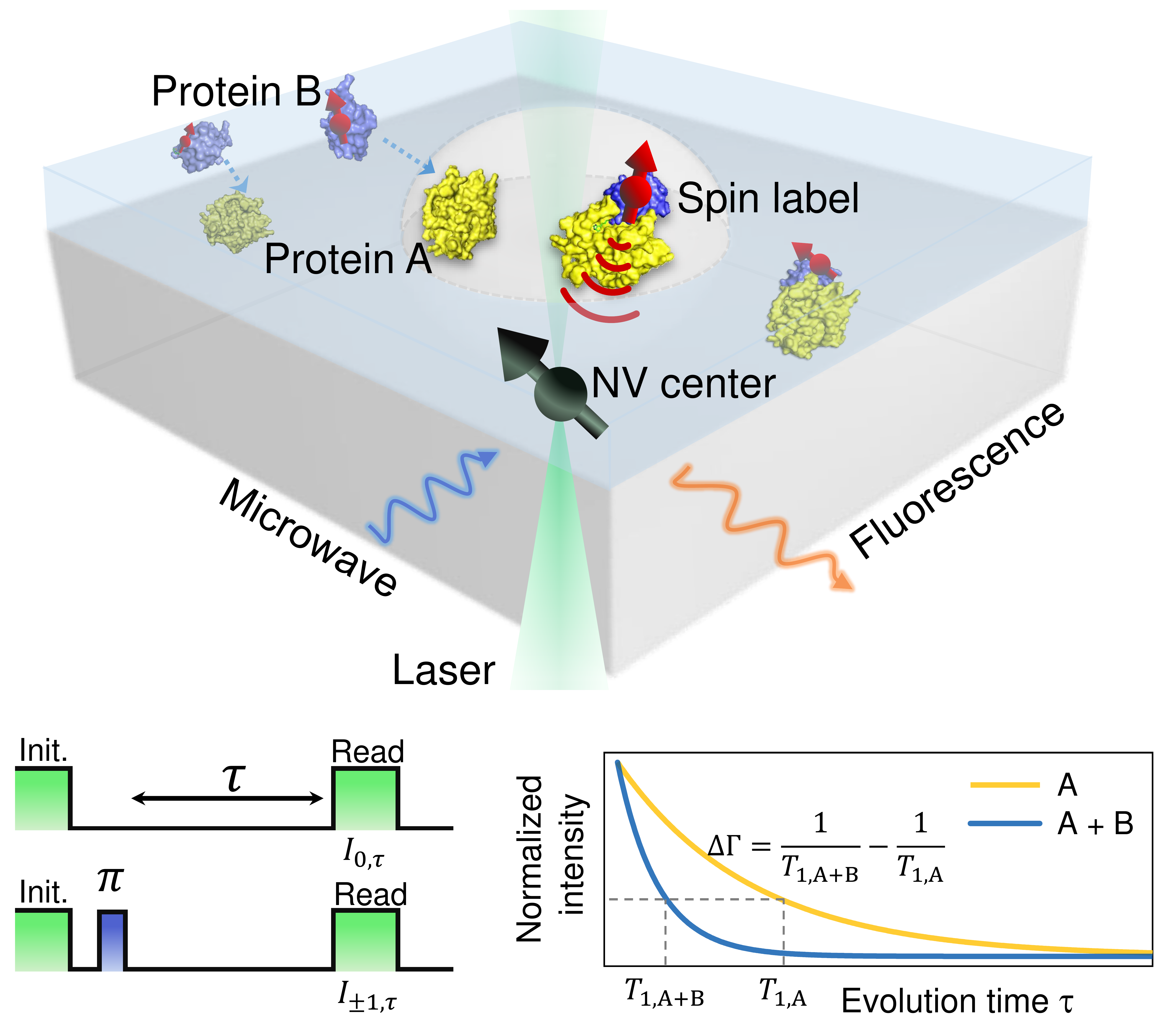}
    \put(-3,25.5){\textbf{B}}
    \put(-1,84){\textbf{A}}
    \put(42,25.5){\textbf{C}}
\end{overpic}
\caption{%\linespread{1}\selectfont
\label{fig:principle} \textbf{Quantum relaxometry for detecting biomolecular interactions using NV centers.} (A) Schematic of the experimental setup. Protein A (yellow) is immobilized on the diamond surface to capture spin-labeled Protein B (blue). The NV center detects magnetic noise from the spin label via acceleration of its spin relaxation rate. (B) $T_1$ measurement sequence. The spin state of the NV center is initialized and read out using 532 nm laser pulses (green). A microwave $\pi$-pulse (blue) is applied to flip the spin state between $|0\rangle \leftrightarrow |\pm1\rangle$. Varying the dark evolution time $\tau$ yields the $T_1$ relaxation profile. (C) $T_1$ curves measured with only protein A (yellow) and with captured spin-labeled protein B (blue). Two relaxation curves were measured using the upper and lower pulse sequences shown in (B), corresponding to initial spin states $|0\rangle$, $|\pm1\rangle$. The difference of the curves was normalized to the reference point at $\tau=0$, yielding the normalized intensity defined as: $(I_{0,\tau}-I_{\pm1,\tau})/(I_{0,0\ \upmu\text{s}}-I_{\pm1,0\ \upmu\text{s}})$.}
\end{figure*}

\noindent\textbf{Principle and conceptual design.} 
The basic idea of our experiment is depicted in \autoref{fig:principle}A. Protein A (yellow) is immobilized on the functionalized diamond surface and then the solution of protein B (blue) labeled with paramagnetic metal spin is introduced onto the interface. If there is an interaction between protein A and B, the capture of B by A results in a close distance between the spin labels and shallow NV centers, thus fluctuating magnetic fields with non-zero root mean square (RMS) of the spin labels will influence NVs' spin relaxation process. Therefore, quantum relaxometry based on the NV center can be used to detect the magnetic noise of the metal-based spin label. The measurement sequence for the relaxation time $T_1$ is performed as \autoref{fig:principle}B. The spin state of NV is polarized to $|0\rangle$ by a 532 nm laser pulse (or to $|\pm1\rangle$ by a following microwave $\pi$-pulse), and freely decays to the thermal equilibrium state after a dark evolution time $\tau$. The difference of decay curves with initial spin state of $|0\rangle$ and  $|\pm1\rangle$ is normalized and the relaxation time $T_1$ is obtained % with an exponential decay fitting 
(\autoref{fig:principle}C). The signal of the metal-based spin label is quantified by the acceleration $\Delta\Gamma$ of the relaxation rate $\Gamma_1=1/T_1$.

In this work, we used Gd$^{3+}$, Mn$^{2+}$ and Cu$^{2+}$ ions chelated by a ligand as the magnetic labels, which are important labels of traditional electron paramagnetic resonance (ESR). They have a long lifetime under laser irradiation, thus providing a sufficient time window for long-term monitoring. We evaluated the photostability of the three labels with a typical laser condition of 30 $\upmu\si{W}$ 532 nm on a home-built optically detected magnetic resonance (ODMR)
spectroscope (\textcolor{blue}{\textit{SI Appendix} Sec. 1}). The results show that the signals of Gd$^{3+}$ and Mn$^{2+}$ are maintained above 60$\%$ after more than 15 h of continuous laser irradiation (\textcolor{blue}{Figure S1}), while the lifetime of Cu$^{2+}$ is $\sim$11 h (\textcolor{blue}{Figure S2}). Therefore, continuous signal accumulation for $\leq$120 h can be performed at room temperature, considering that the laser duty cycle is about 0.1 in most quantum relaxometry experiments. The biotin-ubiquitin (biotin-Ub) complex is a powerful synthetic tool. Combined with the high affinity of biotin to streptavidin (SA), it is widely used in biochemical and cellular research to study ubiquitination---a critical post-translational modification that regulates protein degradation, trafficking, and signaling\cite{bioUb2017scirep,bioUb2024natcomm}. Therefore we chose the interaction between SA and the Mn$^{2+}$ labeled biotin-ubiquitin (biotin-Ub(Mn), details in \textcolor{blue}{Figure S3}) to demonstrate our biomolecular interaction detection approach. 
\\

\begin{figure*}[bp]
\centering
\begin{overpic}[width=.85\textwidth]{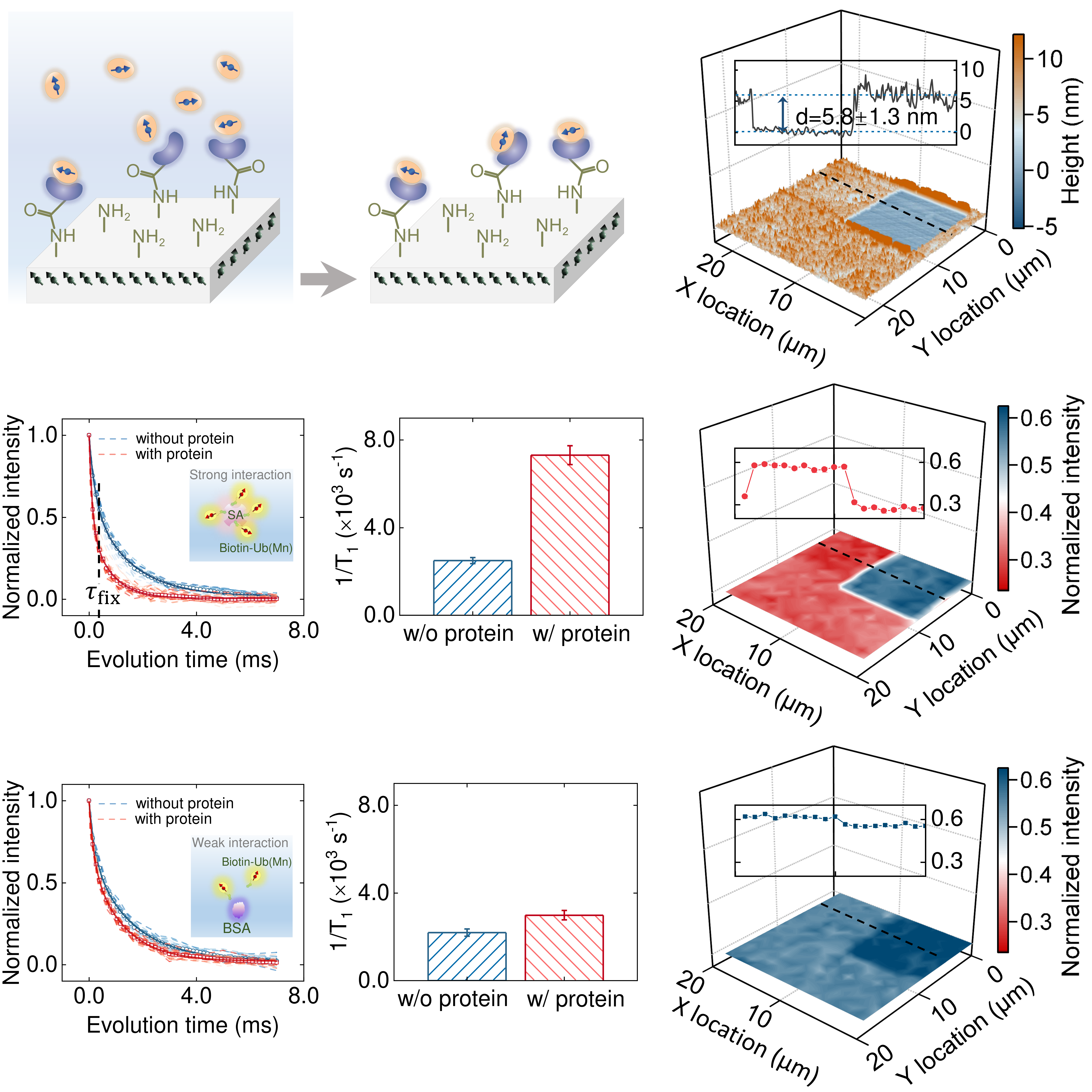}
    \put(0,63){\textbf{C}}
    \put(32,63){\textbf{D}}
    \put(62,63){\textbf{E}}
    \put(0,99){\textbf{A}}
    \put(0,30){\textbf{F}}
    \put(32,30){\textbf{G}}
    \put(62,30){\textbf{H}}
    \put(62,99){\textbf{B}}
\end{overpic}
\caption{\linespread{1}\selectfont
\label{fig:ensemble} \textbf{Micrometer-scale biomolecular interaction detection using ensemble NV centers.} (A) Schematic of biomolecular interaction detection with ensemble NV centers in bulk diamond. The diamond surface is functionalized with a polyethylenimine (PEI) nanogel to achieve surface amination. Then the immobilized proteins, such as streptavidin (SA) or bovine serum albumin (BSA), are sequentially bonded to the surface. Finally, spin-labeled molecules are introduced for detection. (B) Topography image of the diamond surface coated with SA and biotin-Ub(Mn) using atomic force microscope (AFM). Proteins in the square depression area were removed using AFM contact mode. The inset shows a total thickness of 5.8$\pm$1.3 nm. (C-E) Relaxation results of SA+biotin-Ub(Mn), as shown in the inset in (C). (C) $T_1$ curves measured in regions with (27 red lines)/without (21 blue lines) proteins shown in (B). The dashed lines are experimental curves and the hollow dots are averaged data. These dashed lines were fitted using a biexponential decay function to obtain weighted relaxation rates (details in \textit{Materials and Methods}). (D) Comparison of the averaged relaxation rate derived from the dashed lines in (C). The error bar is the standard deviation. The difference in relaxation rates between the two regions is (4.8$\pm$0.5)$\times10^3$ s$^{-1}$. (E) 2D map of the relaxation signal. By fixing the waiting time of the $T_1$ measurement sequence at $\tau_\text{fix}=350\ \upmu\si{s}$, the fluorescence intensity normalized to $\tau=0\ \upmu\si{s}$ was obtained around the depression area shown in (B). \textit{Inset}: Profile of the black line in the 2D map. (F-H) Relaxation results of BSA+biotin-Ub(Mn), as shown in the inset in (F). (F) $T_1$ curves measured in regions with (29 red lines)/without (31 blue lines) proteins. (G) Comparison of the averaged relaxation rate derived from the dashed lines in (F). The relaxation rate difference between the two regions is (0.8$\pm$0.3)$\times10^3$ s$^{-1}$. (H) 2D map of the relaxation signal at $\tau_\text{fix}=350\ \upmu\si{s}$. The proteins in the square depression area have been removed.}
\end{figure*}
%\linespread{1}\selectfont

\noindent\textbf{Diamond surface functionalization.} As an interface analysis method, the experiments required protein immobilization on the surface of the bulk diamond. Commonly used oxygen terminal diamond surface has functional groups such as carboxyl, hydroxyl, carbonyl and ether, in which carboxyl groups can be directly utilized for molecular immobilization\cite{surfaceFunc2022JMCC}. Strategies based on carboxyl residuals have achieved an average molecular spacing of $\sim$20 nm\cite{nanolett2014Gdsingle,nuclear2016science,DNA2018kfnm}. However, interaction studies demand further surface optimization to enhance bonding density while preserving biomolecular activity. We observed that the function of SA was limited by steric hindrance when it was directly attached onto the oxidized diamond surface. Specifically, SA failed to capture biotin-Ub when anchored to tri-acid-oxidized surfaces (\textcolor{blue}{Figure S4}, \textcolor{blue}{Figure S5}). To address this, we used PEI as a buffer layer to increase the distance between SA and the diamond surface. PEI, a polymer rich in amine groups, has shown significant effectiveness for the functionalization of nanodiamonds with small molecules\cite{funcdiamond2024radical,Wuyk2021PEI,nanolett2021PEI}. Here, we extended this strategy to the application for protein conjugation on the surface of bulk diamond, demonstrating both high immobilization efficiency and preserved protein biological activity. The bulk diamond surface is carboxylated with a tri-acid solution, and then an amino cross-linked thin layer is formed on the diamond surface with the mixed solution of PEI and 4 arm-PEG-NHS used to form an amino cross-linked thin layer on the diamond surface (details in \textcolor{blue}{\textit{SI Appendix} Sec. 3.2}). The freshly prepared PEI exhibits a thickness of 1-6 nm and it can collapse to $\ll$1 nm after being stored under room temperature atmospheric conditions for extended periods exceeding 24 h. Notably, the PEI layer showed no measurable impact on the $T_1$ relaxation of NV centers (\textcolor{blue}{Figure S6}). Additionally, the density and uniformity of the Ub(Mn) bonding on the PEI surface are evaluated by spin signals. Spin signal analysis revealed an average Ub(Mn) bonding distance of $\sim$10 nm (\textcolor{blue}{Figure S7}), with only $\sim13\%$ signal variation across four regions spaced 500 $\upmu\si{m}$ apart, confirming high surface homogeneity (\textcolor{blue}{Figure S8}). {Besides, fluorescence analysis also shows a consistent bonding spacing of $\sim$8 nm (\textcolor{blue}{Figure S9}).} Subsequent interaction experiments demonstrated that the PEI-functionalized surface effectively mitigates steric hindrance, providing a robust platform for biomolecular interaction studies.\\

\noindent\textbf{Micrometer-scale detection of biomolecular interaction with ensemble NV centers.} 
Initially, we employed a bulk diamond containing shallow ensemble NV centers, with an average depth of 6.5 nm below the [100] surface, to demonstrate the reliability of the method in studying the overall behavior of molecular interactions. As illustrated in \autoref{fig:ensemble}A, the diamond surface was functionalized with PEI to introduce amino groups, followed by the immobilization of SA or BSA (bovine serum albumin) onto the surface. Subsequently, a solution of biotin-Ub(Mn) was added to the diamond and allowed to react for 40 minutes. The diamond was then cleaned with deionized water and dried using nitrogen gas. We utilized the atomic force microscope (AFM) to characterize the diamond surface after the bonding process, and the result for SA+biotin-Ub(Mn) is presented in \autoref{fig:ensemble}B. AFM contact mode was employed to remove the protein from a square area and the surrounding topography was scanned using tapping mode of AFM. The measured height of the protein layer is 5.8$\pm$1.3 nm. Given that the average hydrodynamic diameters of SA and Ub are $\sim$5 nm\cite{SAsize2008} and $\sim$3 nm\cite{Ubsize1987}, the 5.8-nm height suggests the formation of a protein monolayer. 

\autoref{fig:ensemble}C illustrates $T_1$ curves of the region in \autoref{fig:ensemble}B with (red) and without (blue) protein coating. This comparison highlights a significant relaxation acceleration for the protein-coated region relative to regions without proteins or coated only with SA (as depicted in \textcolor{blue}{Figure S10}). To quantify the signal intensity, we fitted the $T_1$ curves using a biexponential decay function $y=A_\text{short}\exp(-t/T_{1,\text{short}})+A_\text{long}\exp(-t/T_{1,\text{long}})$, which provided superior fitting compared to the single-exponential decay, consistent with prior studies\cite{sciadv2021radical,redox2022radical,acscent2023radical,nanolett2013Mn}. These studies have primarily emphasized the slow relaxation component (considered more sensitive to external spin signals) while neglecting the fast relaxation component from surface-proximal NV centers. Our Monte Carlo simulations reveal that shallow NV centers contribute a higher proportion to the detectable signal (details in \textit{Materials and Method}), as the relaxation change 
$\Delta\Gamma$ originating from the coupling strength between NV and spin label exhibits a stronger sixth-power inverse dependence on separation distance $r$ ($\Delta\Gamma\propto1/r^6$, \textcolor{blue}{\textit{SI Appendix} Sec. 8.2}), which outweighs the fourth-power depth dependence of NV relaxation rate $\Gamma_1$ on depth $d$ ($\Gamma_1\propto1/d^4$)\cite{PRL2014myerssurface,PRL2014surfacenoise}. To utilize this fast component, the ensemble relaxation rate was quantified using the amplitude-weighted characteristic rate $\Gamma_{1,\text{w}}=(A_\text{short}/T_{1,\text{short}}+A_\text{long}/T_{1,\text{long}})/(A_\text{short}+A_\text{long})$, yielding a sensitivity enhancement of up to 4.5-fold (details in \textit{Materials and Methods}). As shown in \autoref{fig:ensemble}D, the mean $\Gamma_{1,\text{w}}$ for protein-coated regions (red) exceeded that of bare regions (blue) by (4.8$\pm$0.5)$\times10^3$ s$^{-1}$. {This fast relaxation acceleration indicates that SA has a high binding efficiency for the biotin-Ub complex, which is attributed to the high affinity between SA and the biotin component of the complex.} Then we fixed the dark evolution time $\tau$ of $T_1$ measurement sequence and scanned around the square region without protein. The relaxation rate distribution was visualized through the normalized fluorescence intensity. The region with protein exhibited faster relaxation, corresponding to lower fluorescence intensity. The 2D map in \autoref{fig:ensemble}E aligns with the AFM topography. In addition, we substituted SA with BSA{, which is commonly used as a blocking agent and lacks specific biotin-binding sites like SA (inset in \autoref{fig:ensemble}F). However, the surface plasmon resonance (SPR) analysis revealed weaker binding efficiency between BSA and biotin-Ub complex (\textcolor{blue}{Figure S11}). This observation was corroborated by our relaxation measurements.} {Similarly, the AFM topography indicates a protein monolayer with a thickness of $\sim8$ nm (\textcolor{blue}{Figure S13}).} The relaxation rates in regions with/without proteins show a little difference and the mean relaxation acceleration is (0.8$\pm$0.3)$\times10^3$ s$^{-1}$ (\autoref{fig:ensemble}G-H). {In addition, we demonstrated the method’s efficacy under solution conditions (\textcolor{blue}{Figure S22}).}
The ensemble NV-based results highlight the capability of our quantum relaxation methodology in effective differentiation between strong and weak biomolecular interactions, holding potential applications for molecular recognition and screening. 
\\

\begin{figure*}[bp]
\centering
\begin{overpic}[width=.85\textwidth]{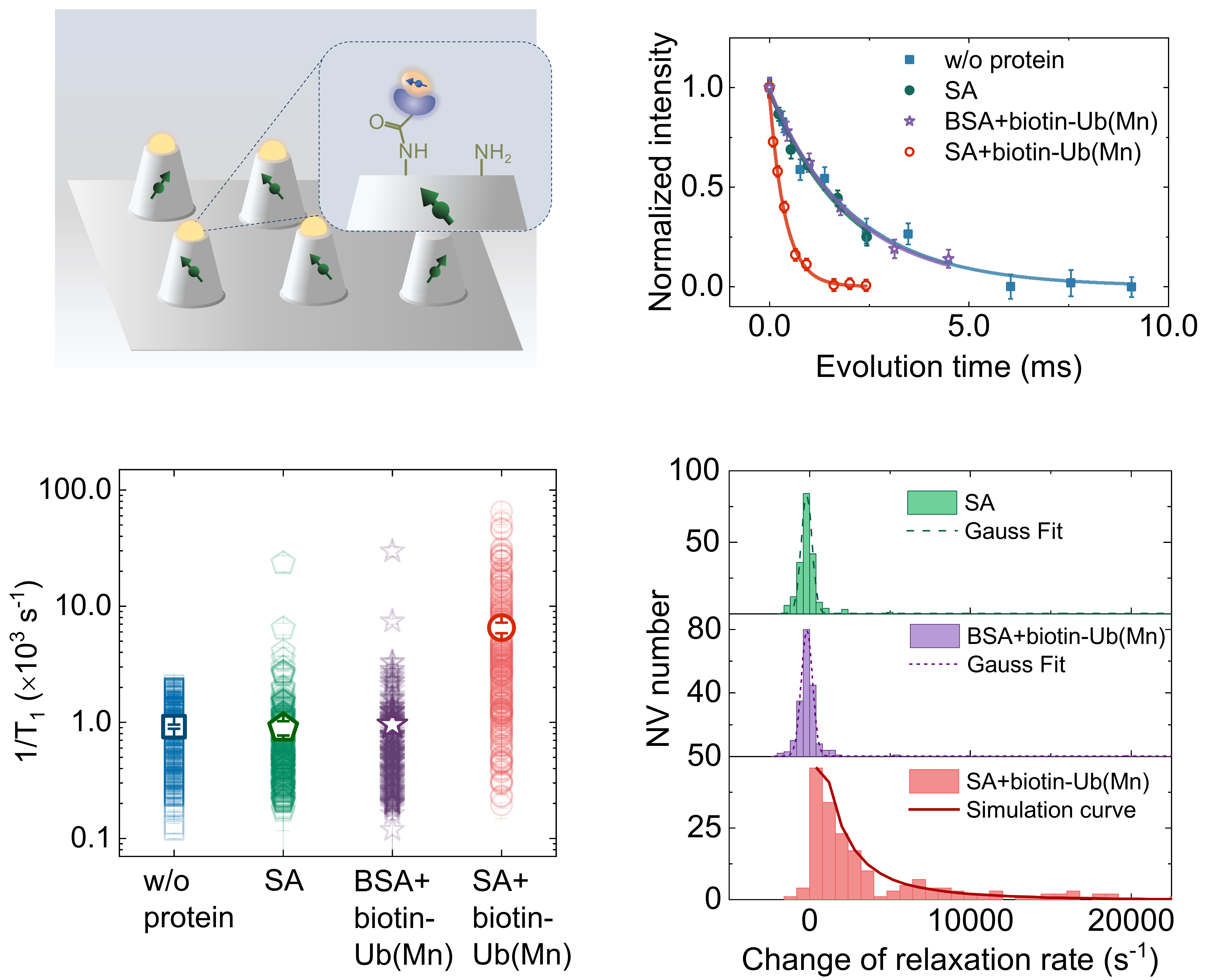}
    \put(1,43){\textbf{C}}
    \put(1,80){\textbf{A}}
    \put(52,43){\textbf{D}}
    \put(52,80){\textbf{B}}
\end{overpic}
\caption{\linespread{1}\selectfont
\label{fig:pillarinteraction} \textbf{Nanoscale biomolecular interaction detection using single NV centers.} (A) Schematic of the interaction detection with single NV centers in nano-pillar. The expanded view shows the side profile of a nano-pillar coated with PEI, pre-immobilized proteins (e.g., SA or BSA), and spin-labeled protein (e.g., biotin-Ub(Mn)). (B) $T_1$ curves of a typical single NV center under four conditions. The dots are experimental data points and the solid lines are fitting curves to the single exponential decay function. (C) Relaxation rates of 195 single NV centers under these four surface conditions. The mean rates for each condition are presented as mean $\pm$ standard deviation, which are $(0.92\pm0.04)\times10^3\ \si{s}^{-1}$, $(0.9\pm0.1)\times10^3\ \si{s}^{-1}$, $(1.0\pm0.2)\times10^3\ \si{s}^{-1}$, $(6.5\pm0.7)\times10^3\ \si{s}^{-1}$. (D) Histogram distributions of relaxation accelerations for 195 single NV centers relative to the bare diamond. The dashed lines represent Gaussian fits. The fitting results presented as the peak $\pm$ standard deviation are $(-0.2\pm0.3)\times10^3\ \si{s}^{-1}$ (green line, SA) and $(-0.2\pm0.4)\times10^3\ \si{s}^{-1}$ (purple line, BSA+biotin-Ub). The solid red line is the theoretical curve obtained from the Monte Carlo simulation, assuming an average distance of 12 nm between SA proteins (details in \textcolor{blue}{Figure S21}). }
\end{figure*}
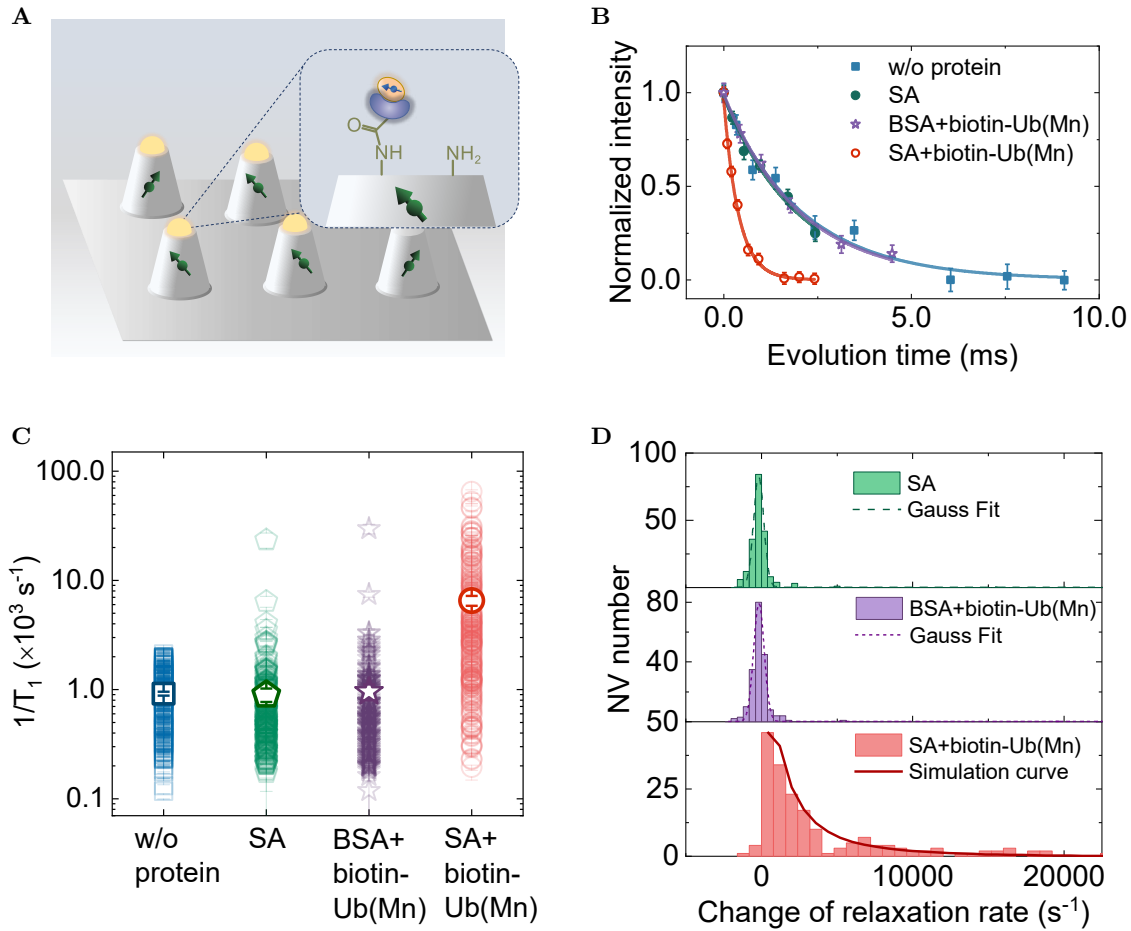
\linespread{1}\selectfont

\noindent\textbf{Nanoscale detection of biomolecular interaction with single NV centers.} 
We next conducted experiments employed single NV centers. The diamond sample was fabricated with nanopillar arrays on the surface and the NV centers were implanted at an average depth of 5.5 nm (details in \textit{Materials and Methods}, \textcolor{blue}{\textit{SI Appendix} Sec. 7.2}). The chemical treatment of the diamond surface followed the same protocol as for ensemble NV diamonds (\autoref{fig:pillarinteraction}A). A total of 195 single NV centers with long background relaxation times ($T_{1,\text{BG}}>$0.5 ms) were selected as quantum sensors. \autoref{fig:pillarinteraction}B shows the $T_1$ curves of a typical single NV center under four conditions: (1) bare diamond, (2) surface bonded only with SA, (3) surface bonded with BSA followed by biotin-Ub(Mn) solution, (4) surface bonded with SA followed by biotin-Ub(Mn) solution. The data points were fitted with a single-exponential decay function. The obtained relaxation rate showed minimal variation under the first three conditions, with significant acceleration observed in the presence of SA + biotin-Ub(Mn). Statistical analysis was conducted on the relaxation rates of the 195 single NV centers. As shown in \autoref{fig:pillarinteraction}C, the average relaxation rates of the four conditions are $(0.92\pm0.04)\times10^3\ \si{s}^{-1}$, $(0.9\pm0.1)\times10^3\ \si{s}^{-1}$, $(1.0\pm0.2)\times10^3\ \si{s}^{-1}$, $(6.5\pm0.7)\times10^3\ \si{s}^{-1}$. The change of average relaxation rate for BSA + biotin-Ub relative to the bare diamond ($\sim0.08\times10^3\ \si{s}^{-1}$) is smaller than that observed in the above ensemble NV experiments, which could potentially be attributed to the selection of relatively deep single NV centers (background rate, i.e. without protein, $\Gamma_{1,\text{BG}}<2000\ \si{s}^{-1}$) and the variations in surface modifications observed between diamond samples. And the relaxation rate change $\Delta\Gamma=\Gamma_{1, \text{w/ protein}}-\Gamma_{1, \text{BG}}$ for each NV center is given as a histogram in \autoref{fig:pillarinteraction}D.
For the SA + biotin-Ub condition, the distribution of $\Delta\Gamma$ is broad, primarily attributed to the random positioning of proteins and the varying depths of NV centers. It is noted that some NV centers exhibit significantly accelerated relaxation rates induced by protein aggregation (above $1\times10^4\ \si{s}^{-1}$), thereby increasing the average signal intensity beyond typical measurement ranges. Monte Carlo simulations suggest that the mean spacing between SA proteins is approximately 12 nm (details in \textcolor{blue}{\textit{SI Appendix} Sec. 10.2}).
\\

\noindent\textbf{Theoretical simulation for the magnetic signal of metal-based spin label.} From the distribution of relaxation rate acceleration detected by single NV centers, we were able to estimate the number of proteins around NV centers. The theoretical model is employed as below. Generally, the relaxation rate of the NV center is affected by the magnetic noise of its environment and given by
\begin{equation}
    \Gamma_1=\Gamma_{1,\text{BG}}+\Delta\Gamma=\frac{1}{T_{1,\text{BG}}}+3\gamma_e^2\langle B_\bot^2\rangle\frac{\tau_c}{1+\omega_0^2\tau_c^2},
\end{equation}
where $\Gamma_{1,\text{BG}}=1/T_{1,\text{BG}}$ is the background relaxation rate (without proteins), which is due to the lattice relaxation and spin bath near the NV center\cite{PRL2014myerssurface,PRL2014surfacenoise}. Here we simplify the background relaxation rate $\Gamma_{1,\text{BG}}=\Gamma_{1,\text{surf}}+\Gamma_{1,\text{bulk}}$. $\Gamma_{1,\text{surf}}$ is attributed to the magnetic noise of surface impurities, which varies with the depth of NV centers, while $\Gamma_{1,\text{bulk}}$ is independent of NVs' depth (details in \textcolor{blue}{\textit{SI Appendix} Sec. 9.2}). $\gamma_e$ is the NV gyromagnetic ratio. $B_\bot$ is the RMS magnetic field of the metal-based spin labels at the location of the NV center, which is determined by both the protein bonding density and the NV depth. As the width of the spectral density function of Mn$^{2+}$ is $\sim$GHz\cite{nanolett2013Mn}, our experiment is performed at zero magnetic field. Thus the resonant frequency is $\omega_0=2\pi D_\text{gs}$ ($D_\text{gs}=2.87\ $GHz is the zero-field splitting of the NV center between $|0\rangle$ and  $|\pm1\rangle$) and $\tau_c$ is the auto-correlation time of the magnetic noise of spin labels. $\Delta\Gamma$ quantifies the signal intensity, representing the acceleration of the relaxation rate caused by molecular interaction events (details in \textcolor{blue}{\textit{SI Appendix} Sec. 8.2}).

\begin{figure*}[htbp]
\centering
\begin{overpic}[width=.93\textwidth]{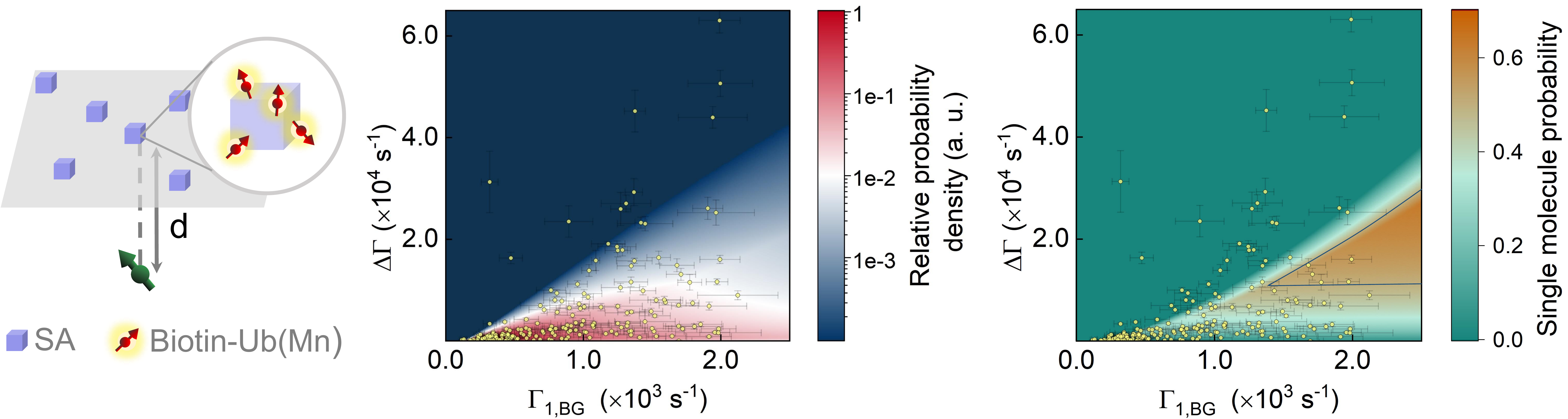}
    \put(-3,27){\textbf{A}}
    \put(25,27){\textbf{B}}
    \put(63,27){\textbf{C}}
\end{overpic}
\caption{\linespread{1}\selectfont
\label{fig:pillar2D} \textbf{Simulation of the magnetic noise signal measured with single NV centers.} (A) Schematic of the Monte Carlo simulation. Single NV centers are modeled with depths ($d$) following a Gaussian distribution ($\mu=5.5\ \si{nm}$, $\sigma=2.8\ \si{nm}$) derived from SRIM ion implantation simulations. SA is approximated as a cubic structure with 4 biotin-binding sites distributed on the diamond surface at an average spacing of 12 nm. (B) Correlation between the NV center's background relaxation rate $\Gamma_{1,\text{BG}}$ and the detected magnetic signal intensity $\Delta\Gamma$. Yellow circles are the experimental data from 195 single NV centers. The colormap displays the simulated probability density of obtaining $\Delta\Gamma$ %the signal 
at given $\Gamma_{1,\text{BG}}$. Outliers (low $\Gamma_{1,\text{BG}}$, high $\Delta\Gamma$) indicate localized protein aggregates or depth estimation deviations. (C) Single-molecule detection probability. The background illustrates the simulated single-molecule detection probability $P_\text{single}$, depicted by the color scale. $P_\text{single}$ quantifies the probability that the observed $\Delta\Gamma$ originates from individual SA-biotin-Ub(Mn) complexes at the given $\Gamma_{1,\text{BG}}$. 
The blue curve marks the contour of $P_\text{single}=0.5$.  
}
\end{figure*}
\linespread{1}\selectfont

To analyze the distribution of relaxation rates, we implemented a Monte Carlo simulation involving 10,000 single NV centers, where the depth of NV centers approximately follows a Gaussian distribution according to the Stopping and Range of Ions in Matter (SRIM) simulation (\textcolor{blue}{Figure S18}C). First, we derived the density of the surface paramagnetic impurities from the statistical distribution of $\Gamma_{1,\text{BG}}$. It is then assumed that SA proteins, which have an approximate cubic shape with a size of 5.8 nm, are randomly distributed on the diamond surface, with each SA capturing four biotin-Ub molecules (\autoref{fig:pillar2D}A). The relaxation acceleration $\Delta\Gamma$ of 10,000 NV centers at a specific SA bonding density was thus simulated and analyzed statistically. By comparing the distribution of the simulated $\Delta\Gamma$ with the experimental data, the average SA bonding density consistent with the experimental results was obtained. The solid red line in \autoref{fig:pillarinteraction}D represents the simulated statistical distribution that best fits the experiment, corresponding to an average SA bonding distance of 12 nm, i.e. a bonding density $\sigma_\text{SA}\approx0.007\ \si{nm}^{-2}$ (\textcolor{blue}{Figure S21}).

We further analyzed the probability of single SA proteins interacting with biotin-Ub(Mn) as detected by single NV centers. \autoref{fig:pillar2D}B shows the relationship between background relaxation rates of NV centers and signal intensity $\Delta\Gamma=\Gamma_{1, \text{SA+biotin-Ub}}-\Gamma_{1, \text{BG}}$ at $\sigma_\text{SA}\approx0.007\ \si{nm}^{-2}$. The color map depicts the probability density derived from Monte Carlo simulations at specified background relaxation rate $\Gamma_{1,\text{BG}}$ and signal intensity $\Delta\Gamma$. Simulations indicate that NV centers with larger background relaxation rates are more likely to measure higher signals. Experimental data points from 195 single NV centers (yellow dots) align closely with the simulated distribution, with the majority of data points clustering in high-probability regions. A small subset detects anomalously high signals (scattered data points in low-probability regions). These outliers can be attributed to two factors: surface protein aggregates generating strong localized magnetic noise, or discrepancies between theoretical and actual NV depths. Specifically for the second, NV centers shallower than predicted---due to sparse local surface electron distribution---exhibit enhanced coupling with spin labels on proteins, causing their signals to deviate from statistical expectations. In our analysis, we defined a single-molecule detection event as one where over 70$\%$ of the signal originates from a single SA-biotin-Ub(Mn) complex, based on Monte Carlo simulations. The probability of detecting such events was mapped in a two-dimensional parameter space of background relaxation rate ($\Gamma_{1,\text{BG}}$) and relaxation acceleration ($\Delta\Gamma$), as shown in \autoref{fig:pillar2D}C. The results indicate that for most of the 195 NV centers, the probability of detecting a single SA-biotin-Ub(Mn) complex is below 0.5, with only 5 NV centers falling within the range of 0.5–0.7. This low probability is attributed to the selection of relatively deep NV centers. To improve detection efficiency, selecting NV centers with shallower depths (2.5–5 nm) is recommended, as their closer proximity to the surface enhances sensitivity to interactions at the single-molecule level.

\section*{Discussion}

In conclusion, we have developed a quantum relaxation-based molecular interaction analysis method utilizing NV centers as quantum sensors and paramagnetic metal spins as labels, enabling multiscale analysis of biomolecular interactions from micrometer to nanoscale levels. For ensemble NV measurements at the micrometer scale, {they achieved superior sensitivity owing to exceptional fluorescence counting rates and large effective sensing areas. Moreover,} we introduced a refined relaxation rate evaluation protocol for ensemble NV centers, achieving an up to 4.5-fold sensitivity enhancement. This improvement facilitate studies of transient biological phenomena with NV relaxometry, particularly \textit{in vivo} free radical dynamics\textemdash a critical frontier in biomedical research \cite{sciadv2021radical,redox2022radical,acscent2023radical,acsnano2023radical,Nanolett2023radical,funcdiamond2024radical,PNAS2024radical}. {However, the spatial resolution of ensemble measurements remains constrained by the diffraction limit ($\sim$600 nm spot diameter), where signals averaged over hundreds of NV centers obscure nanoscale heterogeneity. In contrast, single NV centers provide nanoscale resolution ($<$10 nm), enabling single-molecule detection, which is critical for resolving intrinsic variations in molecular binding kinetics, conformation, and local chemical environments. This nanoscale capability nevertheless faces challenges of experimental efficiency, in which high-throughput pillar arrays and improved surface modification can help.} The PEI nanogel was proven to be effective to provide a universal surface modification strategy with $\sim$0.01 nm$^{-2}$ protein bonding density and reduced steric hindrance. {Moreover, the amide bond linking PEI to the carboxylated diamond surface can be hydrolyzed using alkaline solutions, which permits controlled surface regeneration\cite{Fischer2010SPRprotocol}. Combined with microfluidics, this recyclable interface could enable \textit{in situ} cleaning and re-functionalization, allowing rapid sensor reconfiguration for diverse analytes in diagnostic arrays and enhancing cost-efficiency.} Further optimizations include NV depth control ($<$5 nm) to increase single-molecule detection probability and readout methods selection like spin-to-charge conversion readout schemes\cite{spin2charge2015PRL} to improve experimental efficiency and temporal resolution.

Quantum relaxometry with metal ion spin labels provides a photostable platform for detecting biomolecular interactions, overcoming the photobleaching limitations of conventional optical techniques and enabling long-term monitoring. While Gd$^{3+}$/Mn$^{2+}$ are detected at zero magnetic field due to GHz-scale magnetic noise, Cu$^{2+}$ requires external magnetic field alignment ($\sim$450 Gs) due to hundreds of MHz-scale spectral linewidth, allowing magnetic "bi-color" labeling in complex biological environments. Strategic integration with emerging microdroplet or microfluidic architectures could further transform this platform into a high-sensitivity, high-throughput, and low-volume biodetection system\cite{microfluidic2023natreview,highthrough2024nanopore}, with promising applications in drug screening. Moreover, the demonstrated single-molecule sensitivity opens unprecedented opportunities in clinical diagnostics, particularly for liquid biopsy applications requiring ultralow-abundance biomarker detection\cite{singleMol2020biomarker,biomarker2022natcomm}.  Additionally, combining our relaxometry approach with nanoscale biointerfaces such as nanopillar arrays, offers unique capabilities for \textit{in situ} cellular activity studies\cite{nanoprobe2022review}. This includes probing plasma membrane protein organization\cite{PNAS2019actinreorgnz,nanolett2022proteincluster}, monitoring nuclear deformation\cite{nanopillar2015nucleardeform}, and tracking ion channel dynamics\cite{PNAS2010ionchannel}. The minimally invasive nature of NV centers further permits intracellular biosensing without disrupting cellular integrity. In short, this work not only provides a fundamental analysis of molecular interactions but also establishes a versatile platform based on large-scale quantum biosensor arrays, promoting a crucial step toward the construction of a high-throughput and high-sensitivity platform and the engineering applications of quantum biotechnology. 

\begin{figure*}[bp]
\centering
\begin{overpic}[width=.95\textwidth]{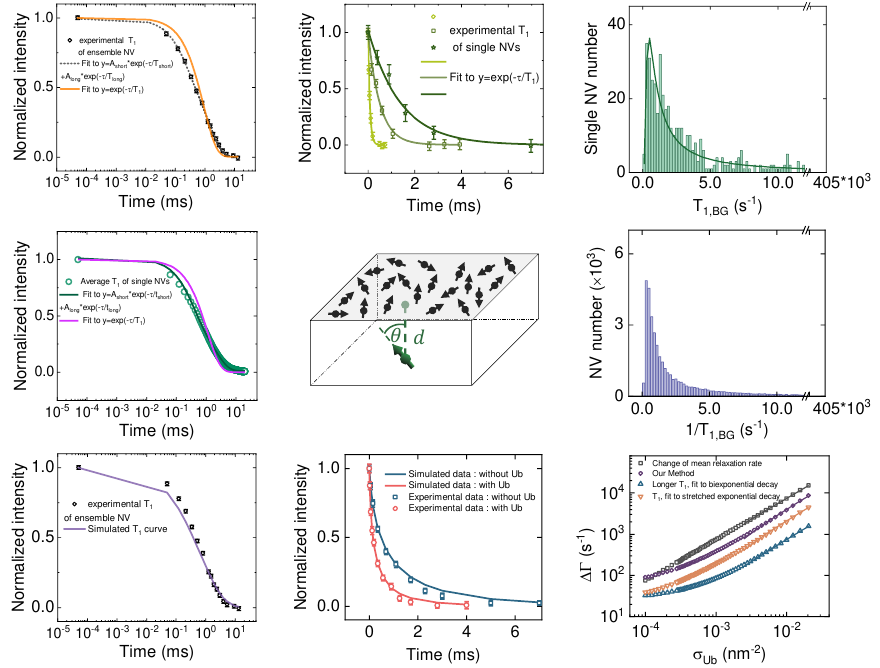}
    \put(0,24){\textbf{G}}
    \put(0,50){\textbf{D}}
    \put(0,76){\textbf{A}}
    \put(32,76){\textbf{B}}
    \put(65,76){\textbf{C}}
    \put(32,50){\textbf{E}}
    \put(65,50){\textbf{F}}
    \put(32,24){\textbf{H}}
    \put(65,24){\textbf{I}}
\end{overpic}
\caption{%\linespread{1}\selectfont
\label{fig:method} \textbf{New evaluation method for relaxation rate of ensemble NV center.} (A) $T_1$ curve of an ensemble NV center. The solid (dashed) line is the (bi)exponential decay fitting curve. (B) $T_1$ curves of 3 single NV centers. Solid lines are fitting curves to the single exponential decay function. (C) Histogram of background relaxation rates of 471 single NV centers. The solid line is a fitting curve based on the surface magnetic noise model (\textcolor{blue}{\textit{SI Appendix} Sec. 9.2}, \textcolor{blue}{Figure S20}). (D) Averaged $T_1$ data of 471 single NV centers in (C). The pink (green) line is a fitting curve to the (bi-)exponential decay function. (E) Schematic of surface magnetic noise model for shallow NV center. Bulk magnetic noise and surface paramagnetic impurities jointly influence background relaxation rates. (F) Simulated distribution of $\Gamma_{1,\text{BG}}$ for shallow single NV centers as $\sigma_\text{surf}=0.40\ \si{nm}^{-2}$ (\textcolor{blue}{\textit{SI Appendix} Sec. 9.3}). (G) Comparison of the simulated $T_1$ curve (purple line) and the experimental $T_1$ data (black dots) for the ensemble NV center. The simulated $T_1$ curve is the average data of single exponential decay curves from the distribution in (F). (H) Comparison of simulated (solid lines) and experimental $T_1$ data (hollow dots). The diamond surface is only bonded with Ub(Mn) and the derived bonding density is $\sigma_\text{Ub}=(1/9)^2\ \si{nm}^{-2}$. (I) Dependence of the ensemble NV relaxation rate change $\Delta \Gamma$ on the Ub(Mn) bonding density $\sigma_\text{Ub}$ with relaxation rate evaluation methods: (1) characteristic rate $1/T_{1,\text{stre}}$ from stretched exponential fits, (2) slower characteristic rate $1/T_{1,\text{long}}$ from biexponential fits, (3) weighted rate $\Gamma_{1,\text{w}}$ (our method). Black points are the arithmetic average of relaxation rate accelerations for single NV centers constituting an ensemble NV center. {The solid lines are linear fitting curves of data corresponding to the bonding spacing from 20 nm to 7 nm.} }
\end{figure*}

\matmethods{
\subsection*{Biexponential shape of $T_1$ curve of ensemble NV centers}
The $T_1$ curve of the shallow ensemble NV center exhibits a biexponential decay, as shown in \autoref{fig:method}A, which has been reported\cite{sciadv2021radical,redox2022radical,acscent2023radical,nanolett2013Mn}. The fast relaxation component is usually attributed to the effect of cross-relaxation of NV pairs, or NV centers close to diamond surface. Given that the implanted dose of the ensemble diamond used in this work is $10^{-13}\ \si{cm}^{-2}$ (with an estimated average distance of $\sim$30 nm between NV centers), the effect of cross-relaxation can be neglected. We excluded the impact of charge relaxation using a 594 nm laser for readout, and observed that the $T_1$ curve of the ensemble NV center still exhibited a biexponential shape (\textcolor{blue}{Figure S16}). According to the following simulations, the biexponential behavior is demonstrated to arise from the heterogeneous relaxation rates of individual NV centers within the ensemble.
 
 Single shallow NV centers typically display single-exponential $T_1$ decay (\autoref{fig:method}B, \textcolor{blue}{Figure S17}), but their relaxation rates vary widely ($100-40000\ \si{s}^{-1}$, \autoref{fig:method}C). Superimposing these 471 single-exponential $T_1$ curves reproduces the ensemble's biexponential profile (\autoref{fig:method}D), confirming the attribution of the biexponential shape to the relaxation rate broadening of single NVs. As described in the main text, the depth-dependent surface magnetic noise broadens the relaxation rates of single NV centers (\autoref{fig:method}E, \textcolor{blue}{\textit{SI Appendix} Sec. 9.2}). According to the SRIM simulation, NV depths follow approximatively a Gaussian distribution (truncated at $\geq2$ nm from the surface considering the shallow NVs' instability; \textcolor{blue}{Figure S18}C). Averaging $T_1$ curves of 40,000 single NV centers with surface spin density of $\sim$0.40 $\si{nm}^{-2}$ simulates the experimental $T_1$ curve of ensemble NV center (\autoref{fig:method}F-G, \textcolor{blue}{Figure S19}), the consistency demonstrateing the reliability of the surface magnetic noise model.\\

\subsection*{Evaluation for relaxation rates of ensemble NV centers}
The biexponential decay reflects contributions from shallower (fast-relaxing) and deeper (slow-relaxing) NV centers. To enhance the sensitivity, we introduce a weighted relaxation rate: $\Gamma_{1,\text{w}}=(A_\text{short}/T_{1,\text{short}}+A_\text{long}/T_{1,\text{long}})/(A_\text{short}+A_\text{long})$, where $A_\text{short}$, $A_\text{long}$ and $T_{1,\text{short}}$, $T_{1,\text{long}}$ are amplitudes and relaxation times from biexponential fitting. This evaluation method is compared to two conventional approaches: (1) using the slower component $T_{1,\text{long}}$ from biexponential fits\cite{sciadv2021radical,redox2022radical,acscent2023radical,nanolett2013Mn}, and (2) extracting $T_{1,\text{stre}}$ from stretched-exponential fits\cite{SciAdv2024cellforce,PRA2024zc,PRL2017ensembleT1,APL2015EnsembleT1,T1PRL2012tempmag}. We presented a comparison of the three methods by numerical simulations as follows: firstly, simulated background $T_1$ curve of ensemble NV center were generated by averaging 40,000 individual decays $y_i=\exp (-t\cdot(\Gamma_{1,i}+\delta\Gamma_{1,i})$. Then local relaxation accelerations $\delta\Gamma_{1,i}$ for each single NVs were introduced due to the randomly distributed Ub(Mn) on the surface at density $\sigma_\text{Ub}$. These modified decays $y_i=\exp (-t\cdot(\Gamma_{1,i}+\delta\Gamma_{1,i})$ were averaged to produce the $T_1$ curve with Ub protein. \autoref{fig:method}H shows a typical comparison of simulated (solid lines) and experimental $T_1$ data (yellow dots) at $
\sigma_\text{Ub}=(1/9)^2\ \si{nm}^{-2}$. The relaxation rate difference between with and without Ub $\Delta\Gamma$ reflects the simulated measured signal intensity. All three methods exhibit linear responses to Ub(Mn) density (\autoref{fig:method}I). However, $\Gamma_{1,\text{w}}$ captures 2.0$\times$ and 5.5$\times$ greater sensitivity to protein bonding than $1/T_{1,\text{long}}$ and $1/T_{1,\text{stre}}$, respectively, as quantified by linear-fit slopes {around the experimental bonding density}. While all methods underestimate the true single-NV average signal (black points, $\sum_i\delta\Gamma_{1,i}/40000$), $\Gamma_{1,\text{w}}$ aligns most closely with the average signal intensity. {Despite nonlinearity at low densities of $10^{-4}-10^{-3}\ \si{nm}^{-2}$, $\Gamma_{1,\text{w}}$ nearly equals the true average signal.}\\

\subsection*{Diamond membrane}
Single crystalline diamond membranes (Element Six, Electronic Grade) were used to create NV centers via ion implantation. Ensemble NV centers were generated by implanting 8 keV $^{14}\text{N}_2^+$ ions at a dose of $1\times10^{13}\ \si{cm}^{-2}$ into the [100] cut bulk diamond (2 mm$\times$2 mm$\times$0.5 mm), followed by vacuum annealing (800 $^\circ$C, 2 h). SRIM simulations indicated an average NV depth of $6.5\pm2.8\ $nm and a density of $\sim1000\ \upmu\si{m}^{-2}$. The uniformity of fluorescence and spin properties is demonstrated in \textcolor{blue}{Figure S14}.

For single NV center experiments, an ultra-thin diamond membrane (2 mm$\times$2 mm$\times$0.05 mm) was used. Four regions of the diamond were implanted with 4 keV $^{15}\text{N}^+$ at doses of $1\times10^{11}\ \si{cm}^{-2}$, $3\times10^{11}\ \si{cm}^{-2}$, $6\times10^{11}\ \si{cm}^{-2}$, $1\times10^{12}\ \si{cm}^{-2}$. Cylindrical nano-pillars were fabricated by a self-aligned patterning technique\cite{wmq2022sciadv}. The process began with spin-coating the diamond surface with double layers: a lower polydimethylglutarimide (PMGI) layer ($\sim270\ $nm) and an upper polymethyl methacrylate (PMMA) layer ($\sim210\ $nm). Electron-beam lithography (EBL) patterned smaller holes in the PMMA, while wet etching with 2.38$\%$ tetramethylammonium hydroxide created larger undercuts in the PMGI layer, forming a dual-mask structure. Nitrogen ions were then implanted through these masks, confining NV center formation to regions within a 15 nm radius. Then the smaller hole array of PMMA on the top was fabricated by electron beam lithography (EBL) and the bigger hole array of PMGI below was generated by the wet etching with 2.38$\%$ tetramethylammonium hydroxide. Next, the N$^+$ ions were implanted into the diamond through the PMMA and PMGI holes. The well-designed double-hole layers constrain the implantation region to within a 15 nm radius from the center. After removing PMMA using acetone, depositing the titanium (Ti) layer by electron beam evaporation, and removing PMGI using N-methyl pyrrolidone, a circular Ti mask array was formed on the diamond surface. The nano-pillars were fabricated by reactive ion etching (RIE) with the mixed gas of CHF$_3$ and O$_2$. The height of the conical cylinder ($\sim350\ $nm) was controlled by the etching time. Finally, the Ti layer was removed with a buffered oxide etch. The diamond was annealed at 1000 $^\circ$C and subjected to a boiling tri-acid solution (98$\%$ H$_2$SO$_4$, 70$\%$ HNO$_3$, 60$\%$ HClO$_4$ at a 1:1:1 volume ratio) for 8 hours to eliminate surface contaminants. A final air-annealing step at 580°C for 20 minutes further optimized NV proximity to the surface, yielding an average depth of $\sim$5.5 nm.

Fluorescence measurements of NV centers within the nano-pillars revealed counting rates of approximately 1.9 Mcps under 532 nm laser saturation (\textcolor{blue}{Figure S15}). We selected the single NV centers with background relaxation times exceeding 500 $\upmu$s as sensors for biomolecular interaction detection to ensure high sensitivity and signal-to-noise ratios.\\

\subsection*{Metal spin-labeled proteins}
In this work, ubiquitins were used to label Gd$^{3+}$, Mn$^{2+}$, or Cu$^{2+}$. The cysteine mutation sites were introduced into Ub and probe molecules 4PhSO2-3Br-PyMTA were linked to these sites (\textcolor{blue}{Figure S3}). Following the chelation reaction with Gd$^{3+}$, Mn$^{2+}$ or Cu$^{2+}$ solutions, the Ub protein can be labeled and Ub(Gd/Mn/Cu) molecules were obtained. The biotin-Ub(Gd/Mn/Cu) complex was formed with the linkage between biotin and Ub(Gd/Mn/Cu) via a short peptide. \\

\subsection*{Diamond surface functionalization with PEI}
The detailed protocol for diamond surface functionalization with PEI is mainly referred to Ref.\citenum{nanolett2021PEI} (as shown in \textcolor{blue}{Figure S6}). (1) Carboxylation of the diamond surface. The diamond surface was carboxylated by immersing it in a tri-acid mixture (H$_2$SO$_4$/HNO$_3$/HClO$_4$) at 180 $^\circ$C for 3 hours. This step was preceded by sequential cleaning in concentrated nitric acid, 1 M NaOH solution and 1 M HCl solution at 100 $^\circ$C for 1 hour each, followed by thorough rinsing with deionized water (ddH$_2$O washing). (2) PEI bonding. The carboxyl groups on the diamond surface were activated with a mixture of 2 mM 1-ethyl-3-(-3-dimethylaminopropyl) (EDC, Thermo Fisher Scientific, 22980) and 5 mM N-hydroxysuccinimide (NHS, Thermo Fisher Scientific, 24510) for 15 minutes. After activation, the diamond was washed to remove the excess EDC/NHS mixture. The activated diamond was then incubated in 66 $\upmu$L of a 15 mg/mL PEI solution (molecular weight$\sim$25 kDa, dissolved in PBS buffer, Sigma-Aldrich, 408727) and allowed to react for 30 minutes at room temperature. (3) PEI cross-linking. To cross-link the PEI layer, 18 $\upmu$L of a 40 mg/mL 4-arm-PEG-NHS solution (molecular weight 20 kDa, dissolved in PBS buffer, Axispharm, AP14802) was added. The mixture was gently shaken for 90 minutes to ensure complete cross-linking. Finally, the diamond was rinsed with ddh$_2$O to remove any unreacted reagents. \\

\subsection*{Protein immobilization on the diamond surface} The streptavidin (SA, Thermo Fisher Scientific, 434301) was prepared at a concentration of 0.6 mg/mL in PBS buffer (pH 7.2), while biotin-Ub(Mn) was dissolved in MES buffer (0.1 M, pH=6.0) at a concentration of 0.6 mM. The PEI-coated diamond was incubated in a freshly prepared mixture containing EDC (20 mg/mL, 1 $\upmu$L), SA (0.6 mg/mL, 5 $\upmu$L) and MES buffer (0.1 M, pH=6.0, 8 $\upmu$L) for 2 hours at room temperature. After the reaction, the diamond was rinsed thoroughly with MES buffer. Following the SA immobilization, 5 $\upmu$L of 0.6 mM biotin-Ub(Mn) was applied to the diamond surface and allowed to react for 40 minutes. The surface was then rinsed repeatedly with MES buffer to remove unbound molecules. The immobilization protocol for bovine serum albumin (BSA) followed the same procedure.
\\

\subsection*{Experimental setup}
The experiments were performed on a home-built confocal microscopy platform comprising the optical, microwave, and control systems. The optical system utilized 532 nm and 594 nm laser beams for spin state initialization and readout. The 532 nm laser (Changchun New Industries, MGL-III-532-150 mW) passed twice through an acousto-optic modulator (AOM, Gooch$\&$Housego, 3200-121), while the 594 nm laser was modulated by a fiber-coupled  AOM (Gooch$\&$Housego, Fiber-Q 633 nm). Both laser beams were combined and delivered through a fiber bundler (Thorlabs, RGB46HF). Then the laser beam was focused on the diamond membrane by an objective (Olympus, PlanApo 60×, numerical aperture (NA), 1.42). The diamond sample was mounted on a high-precision nanopositioner (PI, P562) for scanning. NV center fluorescence was collected through the same objective, filtered by a 561-nm long-pass dichroic mirror (Semrock, Di03-R594-t1-25x36), and detected by avalanche photodiodes (Excelitas, SPCM-AQRH). As for the microwave system, microwaves were generated by a microwave source (Anapico, APSIN6010), and routed through a switch (Mini-Circuits, ZASWA-2-50DRA+). After amplification (Mini-Circuits, ZHL-16-43-S+), the microwaves were delivered to the NV centers via a customized Omega-shaped microwave antenna to manipulate their spin states. An external magnetic field can be applied using a permanent magnet. The system is controlled by an arbitrary sequence generator (Guoyi Quantum (Hefei) Technology, ASG8000) for synchronizing laser and microwave switching, photon data acquisition, and other experimental sequences. Data acquisition and analysis were automated through a dedicated computer, ensuring reproducible experimental workflows.\\

\subsection*{Data Availability}
The AFM image, relaxation data related to the measurements of molecular interactions, and simulation data for ensemble NV centers in this study have been deposited in Zenodo database: https://doi.org/10.5281/zenodo.16780745 \cite{li_dataset}. All other data and materials are included in the article and/or SI Appendix.
}

\showmatmethods{} % Display the Materials and Methods section

\acknow{The authors are grateful to Xingyi Shen, Zhifu Shi and Xiaolan Xu for technical assistance. This work was supported by the National Key R \& D Program of China (Grant Nos. 2019YFA0709300, 2021YFB3202800), the National Natural Science Foundation of China (Grant Nos. T2125011, 12174377), the CAS (Grant Nos. YSBR-068), Innovation Program for Quantum Science and Technology (Grant Nos. 2021ZD0302200, 2021ZD0303204), New Cornerstone Science Foundation through the XPLORER PRIZE, Science and Technology Department of Zhejiang Province (2025C01041) and the Fundamental Research Funds for the Central Universities (226-2024-00142).}

\showacknow{} % Display the acknowledgments section

\bibsplit[31]
%Use \bibsplit to split the references from the body of the text. Value "[2]" represents the number of reference in the left column (Note: Please avoid single column figures & tables on this page.)

% Bibliography
\bibliography{Ref}

\end{document}

\newgeometry{a4paper,top=1in,bottom=0.5in,left=0.8in,right=0.8in}
%\linenumbers
  \maketitle
  \thispagestyle{fancy}
\fancyhead{}
\chead{}
\rhead{}
\lfoot{}
\cfoot{\thepage}   %current page number
\rfoot{}
%\fancyhead[L]{\includegraphics[width=0.15\textwidth]{PNAS_logo.PNG}}
\renewcommand{\headrulewidth}{0pt}
\renewcommand{\footrulewidth}{0pt}

\justifying
{\raggedright\sffamily \textbf{Email: } \href{https://www.ustc.edu.cn/}{zhq2011@ustc.edu.cn}; \href{https://chem.nankai.edu.cn/}{xunchengsu@nankai.edu.cn}; \href{https://www.ustc.edu.cn/}{fzshi@ustc.edu.cn}\\}
\quad \\
\quad \\
{\raggedright\sffamily \textbf{This PDF file includes: }}
Supporting text, \autoref{figs:gdmnstability}-\ref{figs:PEIstability24h}, SI references
\quad \\
\restoregeometry

\tableofcontents
\label{sec:SIcontent}
%\clearpage

\section{\label{sec:stabilityofmngdcu} Photostability of metal spin labels}
Before the experiment of molecular interaction detection, we firstly evaluated the photostability of metal-based spin labels (Gd(III)-, Mn(II)-, and Cu(II)-labeled ubiquitins, Ub(Gd/Mn/Cu)) under laser exposure. Compared to conventional fluorescent labels, metal-based spin labels exhibited significantly enhanced laser resistance, providing an extended time window for long-term monitoring. 

\begin{figure*}[hbt]
    \centering
\includegraphics[width=.9\textwidth]{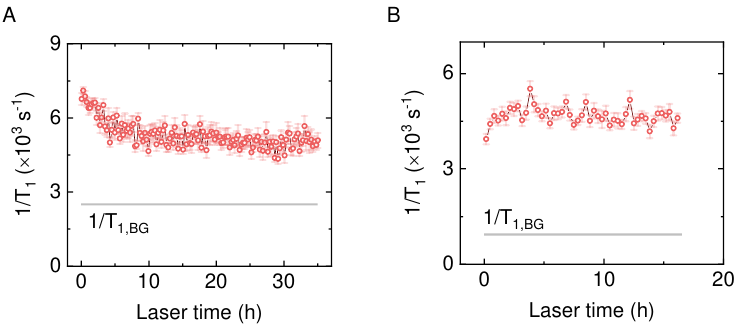}
\caption{\label{figs:gdmnstability} \textbf{Photostability of magnetic signal of Mn(II)- and Gd(III)-labeled ubiquitins.} The red dots are the time trace of relaxation of NVs with (A) Mn(II)- and (B) Gd(III)-labeled ubiquitins. The solid gray lines are the background relaxation rate of the NVs. The measurements were conducted on two different diamond samples. The $x$-axis is the equivalent laser exposure time.
}
\end{figure*}

The photostability of surface-immobilized Gd(III)- and Mn(II)-Ubs was analyzed at zero external magnetic field due to their GHz-level noise spectrum widths (\autoref{figs:gdmnstability}). Under continuous 532 nm laser excitation at 30 $\upmu$W, Mn(II) demonstrated a partial signal attenuation observed within 10 hours while maintaining over 60$\%$ signal retention. In contrast, Gd(III) exhibited exceptional stability, showing no measurable signal loss over the same duration.

\begin{figure*}[htp]
\centering
\includegraphics[width=.9\textwidth]{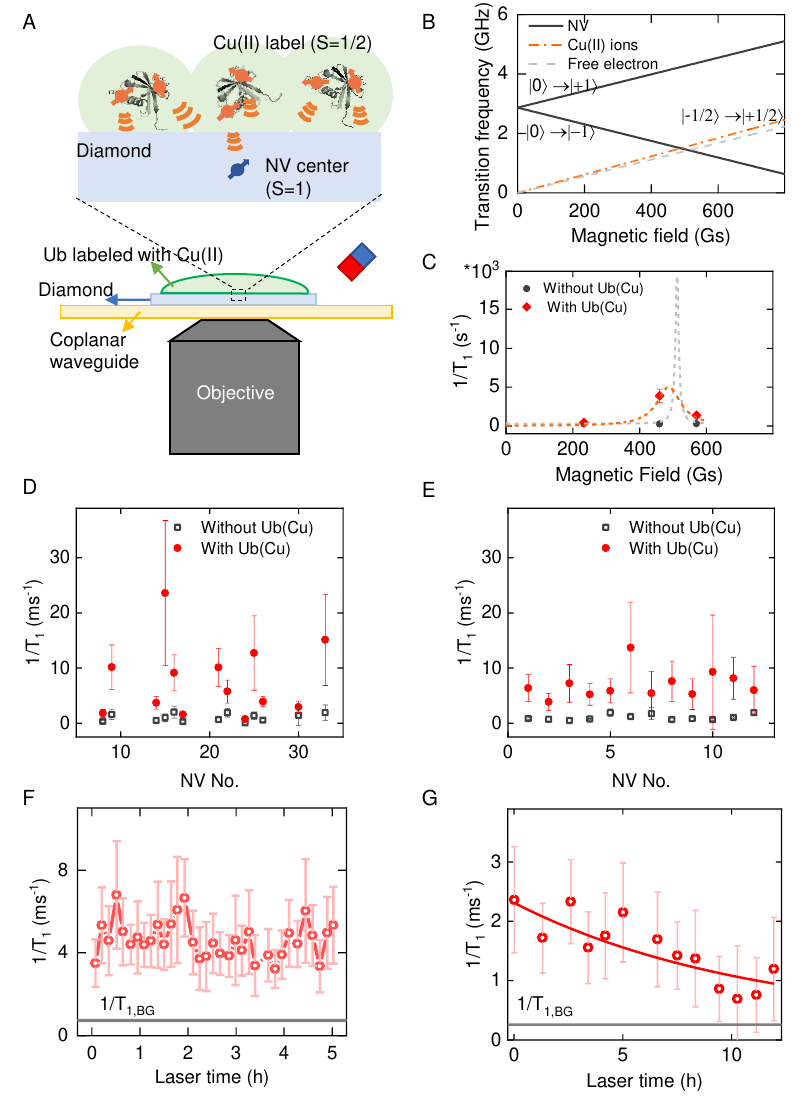}
\singlespacing
\caption{\label{figs:custability} \textbf{Measurement of Cu(II)-labeled ubiquitin.} (A) Experimental schematic. Cu(II) labeled ubiquitin (Ub) is immobilized on the diamond surface. A coplanar waveguide radiates microwaves to manipulate the NV spin state, while an oil-immersion objective focuses the 532 nm excitation laser and collects the fluorescence emitted by the NV center. (B) Dependence of transition frequencies on magnetic field for Cu(II) (red dash-dot line, 487 Gs) ions and free electrons (green dashed line, 512 Gs), differentiated by their distinct $g$-factors (Cu(II): $g_\text{Cu}\approx2.2$; free e$^-$: $g_e\approx2.003$). (C) NV relaxation rates with (red) and without (black) Ub(Cu) at magnetic fields of $B_0=233,\ 466,\ 570\ $Gs. Dashed lines: EPR spectra of Cu(II) (red line) and free electrons (gray line). (D) Single-NV measurements: Mean $\Delta\Gamma=6.8\pm6.3\ \si{ms}^{-1}$ (n=13). (E) Ensemble-NV measurements: Mean $\Delta\Gamma=5.9\pm2.7\ \si{ms}^{-1}$ (n=12). (F-G) Relaxation rate stability under continuous laser exposure. Gray: Background rates. The red line in (G) is the exponential decay fit ($\tau$ = 11 h). $x$-axis is the effective laser time.}
\end{figure*}

For Ub(Cu) analysis (\autoref{figs:custability}A), a Ub(Cu) protein solution was dried onto a diamond surface and detected using a confocal platform under 532 nm laser excitation (30 $\upmu$W power). Given that the noise spectrum width of Cu(II) is on the order of 100 MHz, an external magnetic field of approximately 487 Gs was applied to meet the resonance condition for Cu(II) $\leftrightarrow$ NV interactions. Notably, free electrons near the NV center can be distinguished by resonance magnetic fields due to the distinct $g$ factors (487 Gs for Cu(II) ions ($g_{\text{Cu}}\approx 2.2$) and 512 Gs for free electrons ($g_e\approx2.003$) as shown in \autoref{figs:custability}B)\cite{natcomm2017Cu}. Three magnetic fields ($B_0=233,\ 460,\ 570\ $Gs) were tested to confirm contributions from Cu(II) (\autoref{figs:custability}C). Baseline measurements (black dots, no Ub(Cu)) showed negligible free electron influence at the three fields. In the presence of Ub(Cu), significant NV relaxation acceleration occurred at 460 Gs and 570 Gs. The asymmetry in relaxation rates between magnetic fields of 460 Gs and 570 Gs—symmetrically offset from the free electron resonance (512 Gs)—confirmed Cu(II) as the dominant signal source. Single-NV measurements (\autoref{figs:custability}D) revealed a mean relaxation rate increase of 6.3 ms$^{-1}$ with 6.3 ms$^{-1}$ broadening, attributed to random Ub(Cu) distribution relative to NVs. Ensemble-NV measurements (\autoref{figs:custability}E) yielded a mean rate increase of 5.9 ms$^{-1}$ with 2.7 ms$^{-1}$ broadening, consistent with single-NV statistics. Signal stability under continuous laser exposure was evaluated (\autoref{figs:custability}F-G). The Cu(II) magnetic signal remained stable for 5 hours, and the lifetime is about 11 h. At typical relaxometry duty cycles (0.1), this corresponds to 120 hours of continuous measurement capability at room temperature, demonstrating robust applicability for long-term monitoring.

\clearpage
\section{\label{sec:Ublabeled} Ubiquitin labeled with metal spins}
\begin{figure*}[ht]
    \centering
\includegraphics[width=.75\textwidth]{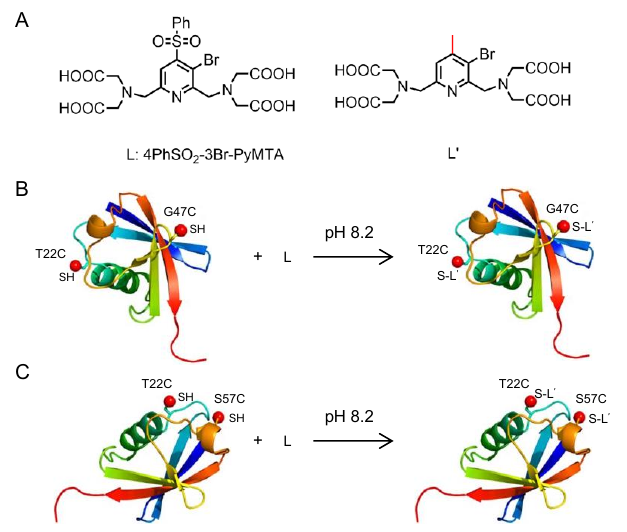}
\caption{\label{figs:Ublabel} \textbf{Schematic of Ub double mutation site linked to the molecule 4PhSO$_2$-3Br-PyMTA.} (A) Structural formula of the linking molecule 4PhSO$_2$-3Br-PyMTA. (B, C) Specific binding of the linking molecule to the double mutation sites T22C/G47C (B) and T22C/S57C (C) of the Ub protein.
 }
\end{figure*}

In our experiments, we used biotin-ubiquitin (biotin-Ub) complex as the target protein. Cysteine mutation sites were introduced into Ub to enable site-specific conjugation with the chelator 4PhSO$_2$-3Br-PyMTA (\autoref{figs:Ublabel}A). \autoref{figs:Ublabel}B-C illustrate two examples of the reaction schematic for 4PhSO$_2$-3Br-PyMTA conjugation to dual cysteine residues on Ub. Subsequent incubation with Gd$^{3+}$, Mn$^{2+}$ or Cu$^{2+}$ solutions allowed metal-specific spin labeling of Ub, forming Ub(Gd/Mn/Cu) complexes. Biotin and Ub were linked with a short peptide. The biotin-Ub(Gd/Mn) complexes averaged four spin labels per complex, compared to two in Ub(Cu). To validate our method, we detected the strong interaction between streptavidin (SA) and biotin-Ub, leveraging the high binding affinity of SA-biotin. Then SA was replaced with bovine serum albumin (BSA) to demonstrate the detection of weak interactions.

\clearpage

\section{\label{sec:diamondfunct} Evaluation of diamond surface functionalization}
\subsection{\label{subsec:diamondCOOH} Carboxy-based functionalization}
\subsubsection{Protein bonding protocol for carboxylated diamond surface}
The diamond surface was oxidized via tri-acid treatment and carboxyl residues enabled covalent conjugation to protein primary amines. Here, the functionalization efficacy was evaluated using metal-based spin signals. The bonding protocol is as follows:

(1) Surface carboxylation. The diamond was immersed in a tri-acid mixture (98$\%$ concentrated sulfuric acid, 70$\%$ concentrated nitric acid and 60$\%$ perchloric acid at a 1:1:1 volume ratio) at 180 $^\circ$C for at least 3 hours to remove organic residues and oxidize the surface. Subsequently, the diamond was sequentially treated with a concentrated nitric acid solution, 1 M NaOH solution and 1 M HCl solutions at 100 $^\circ$C for 1 hour each to enhance carboxyl group density. The diamond was then thoroughly rinsed with ultrapure water to finalize the carboxylation process.

(2) Two-step protein bonding. Carboxyl groups on the diamond surface were activated with 1-ethyl-3-(-3-dimethylaminopropyl) (EDC) and N-hydroxysuccinimide (NHS), forming amine-reactive intermediates. The activated surface was incubated with protein solutions (e.g., Ub(Mn) or SA) for 2 hours at room temperature, enabling covalent immobilization through amide bond formation between surface esters and protein primary amines.

\begin{figure*}[ht]
    \centering
\includegraphics[width=.85\textwidth]{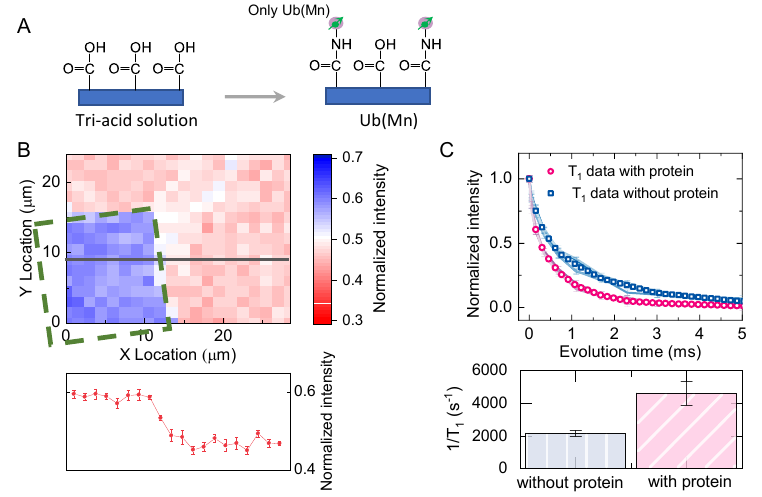}
\caption{\label{figs:cooh-ub} \textbf{Confirmation for relaxation signal of only Ub directly bonded onto the carboxylated surface.} (A) Bonding schematic of Ub(Mn) to the carboxylated diamond surface. (B) Top: 2D map of the normalized fluorescence associated with relaxation rates. $T_1$ measurement sequence is fixed at 2-point waiting dark time $\tau_0 = 0$ and $\tau_1 =350\ \upmu\si{s}$, with initial spin state ${m_s}=0$, $\pm 1$. The pixel values are normalized intensity $(I_{0,\ 350\ \upmu\si{s}}- I_{\pm1,\ 350\ \upmu\si{s}})/(I_{0,\ 0\ \upmu\si{s}}-I_{\pm1,\ 0\ \upmu\si{s}})$ at $\tau_1=350\ \upmu\si{s}$. Ubiquitins within the green dashed box has been removed by AFM contact mode. Bottom: Fluorescence intensity profile along the gray line. High-intensity bottom-left region (no Ub(Mn)) correlates with slower relaxation, consistent with AFM images. (C) Top: $T_1$ curves for regions with (pink, n=3)/without (blue, n=6) Ub(Mn). Bottom: Relaxation rate histogram. Mean signal of Ub(Mn) is $2.5\times10^3\ \si{s}^{-1}$.
 }
\end{figure*}
\subsubsection{Characterization of carboxylated diamond by quantum relaxometry}
Ub(Mn) proteins were covalently bonded to a carboxylated diamond surface via tri-acid treatment (see protocol in \autoref{figs:cooh-ub}A). The surface was dried under $\text{N}_2$(g), and a square region of immobilized proteins was removed via the contact mode of atomic force microscopy (AFM). A 2D relaxation map (\autoref{figs:cooh-ub}B) displays normalized fluorescence intensity at $\tau_1=350\ \upmu\si{s}$ with $T_1$ measurement sequence. The AFM-cleared region (blue square, lower left) exhibits a slower relaxation rate, consistent with the absence of Ub(Mn). $T_1$ curves for regions with (red lines)/without (blue lines) Ub(Mn) were fitted with a biexponential decay function (\autoref{figs:cooh-ub}C). Weighted relaxation rate analysis (details in \textit{Materials and Methods} in main text) revealed a $2.5\times10^3\ \si{s}^{-1}$ acceleration in NV spin relaxation for Ub(Mn)-bound regions, confirming successful surface functionalization.

To evaluate the functionalized surface’s utility for interaction studies, SA was immobilized with the two-step protocol, followed by incubation with biotin-Ub(Gd) solution for 40 min (\autoref{figs:cooh-SA-ub}A). 2D normalized fluorescence intensity map at $\tau_1=950\ \upmu\si{s}$  and $T_1$ curves showed negligible difference between regions without (lower-left in \autoref{figs:cooh-SA-ub}B, removed by AFM) and with proteins (\autoref{figs:cooh-SA-ub}B-C). This suggests that biotin-Ub(Gd) was not captured by SA, which contradicts the SPR results in \autoref{figs:spr}A. The discrepancy may arise from steric hindrance due to the proximity of SA to the diamond surface, potentially impeding biotin-Ub(Gd) binding.
\begin{figure*}[ht]
    \centering
\includegraphics[width=.85\textwidth]{COOH-SA+biotinUb-1.pdf}
\caption{\label{figs:cooh-SA-ub} \textbf{Effect of steric hindrance on biological function of SA directly bonded to carboxylated surface.} (A) Schematic of bonding process: Carboxylated diamond surface$\rightarrow$SA bonding$\rightarrow$biotin-Ub(Gd) capture. (B) Top: 2D map of the signal of relaxation. Proteins within the green box of the lower-left region have been removed by AFM contact mode. Bottom: Fluorescence intensity profile corresponding to the gray line. The fixed evolution time is $\tau_1=950\ \upmu\si{s}$ (methodology as \autoref{figs:cooh-ub}B). (C) Top: $T_1$ curves in regions with (red, n=18) / without (blue, n=18) proteins. Hollow points: average curves. Biexponential fits yield weighted relaxation rates. Bottom: Relaxation rate histogram presented as averages $\pm$ standard deviations. The little difference between regions with and without proteins suggests that there is steric hindrance effect preventing SA from capturing biotin-Ub.
 }
\end{figure*}

\subsection{Diamond surface functionalization with polyethyleneimine (PEI)}\label{subsec:PEIfunct}
\subsubsection{Bonding protocol with PEI}
\begin{figure*}[ht]
    \centering
\includegraphics[width=.85\textwidth]{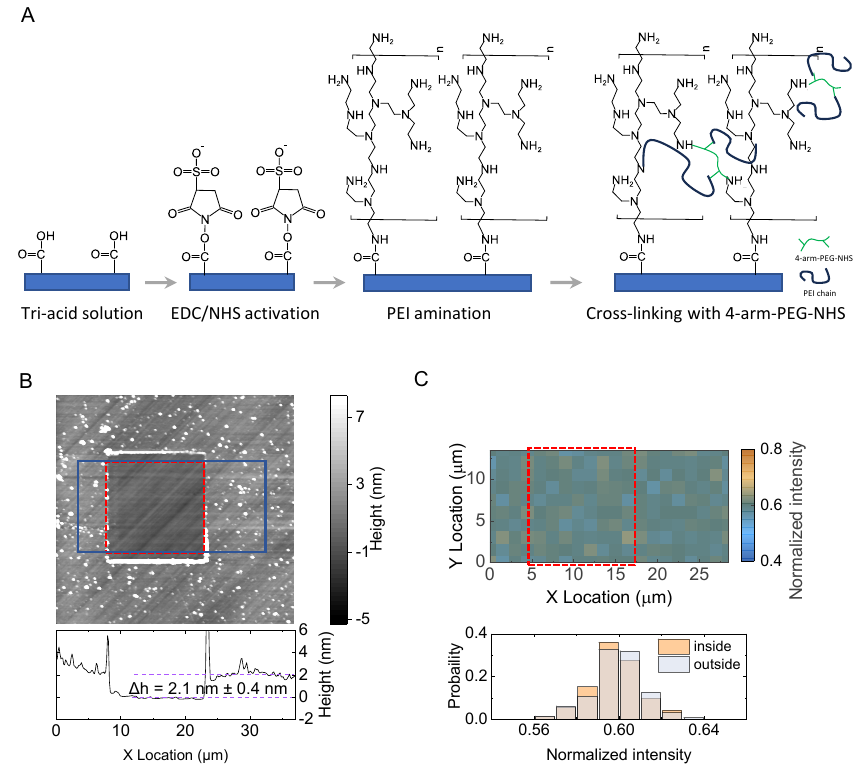}
\caption{\label{figs:PEIbonding} \textbf{Amino functionalization of diamond surface with PEI.} (A) Protocol schematic for diamond surface amino functionalization with PEI. (B) AFM topography image of a PEI nanogel on the diamond surface. PEI within the square depression (marked with a red box) was removed by AFM contact mode. The profile represents the horizontal average curve around the depression, indicating a PEI thickness of 2.1$\pm$0.4 nm. (C) Top: 2D map of the relaxation signal for regions marked with blue box in (B). The pixel is the normalized intensity measured with $T_1$ sequence at the fixed $\tau_1=350\ \upmu\si{s}$. The red box highlights the PEI-free region. Bottom: histogram of normalized intensities inside and outside the red box. Minimal contrast between the inside and outside regions confirms that the PEI nanolayer does not significantly affect NV center relaxation rate, validating its compatibility with quantum sensing applications.
 }
\end{figure*}

To address steric hindrance and achieve high bonding density, we introduced a nanogel layer of amino polymer between the diamond surface and the protein. PEI was chosen as an excellent polymer candidate due to its high amino branching. The detailed bonding protocol is shown in \autoref{figs:PEIbonding}A (also in \textit{Materials and Methods}).

%(1) Diamond surface carboxylation by tri-acid cleaning. 

%(2) PEI bonding. The carboxyl groups on the diamond surface were activated with an EDC and NHS mixture for 15 minutes, followed by washing to remove the mixture. The diamond was then immersed in a PEI solution (66 $\upmu$L of 15 mg/mL PEI with a molecular weight of 25 kDa in PBS buffer) and reacted for 30 minutes at room temperature. Finally, a 4 arm-PEG-NHS solution (18 $\upmu$L of 40 mg/mL with a molecular weight of 20 kDa in PBS buffer) was added and mixed, followed by shaking for 90 minutes. The diamond was then cleaned with ddh$_2$O.

%(3) One-step protein bonding (e.g., SA and Ub(Mn)). The protein solution was mixed with EDC in MES buffer, and the PEI-coated diamond was placed in the mixture for a 2-hour reaction at room temperature. The diamond was then cleaned with MES buffer.

\autoref{figs:PEIbonding}B shows the AFM image of diamond surface coated with a PEI nanogel. The central depression was removed by the contact mode of AFM. The profile line indicates a PEI thickness of 2.1$\pm$0.4 nm. Note that the thickness varies for each different modification, typically ranging from 1-6 nm when freshly prepared, but collapsing to less than 1 nm after being stored at room temperature and atmospheric environment over a day. \autoref{figs:PEIbonding}C scans the normalized fluorescence intensity of the region marked with the blue box in \autoref{figs:PEIbonding}B, measured at the fixed $\tau_1=350\ \upmu\si{s}$ with $T_1$ measurement sequence. The red box marks the region without PEI. The 2D map and histogram show no difference between the inside and outside regions, indicating that PEI does not influence the NV relaxation in dry conditions.

\subsubsection{Density and uniformity evaluation}
\begin{figure*}[ht]
    \centering
\includegraphics[width=.9\textwidth]{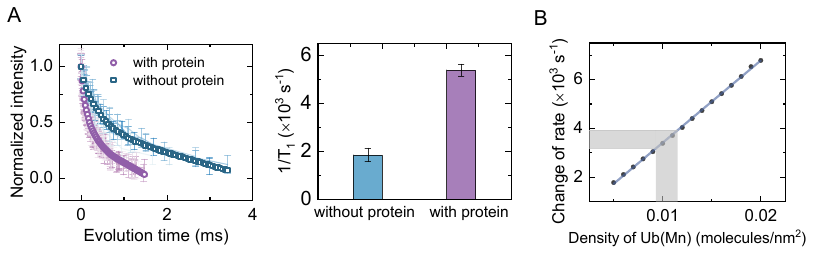}
\caption{\label{figs:PEIdensity} \textbf{Ub bonding density analysis on PEI-coated diamond surface.} (A) Left: $T_1$ curves for regions with (purple data, n=19) and without (blue data, n=6) Ub(Mn). Hollow points: average $T_1$ data in each region. Relaxation rates were obtained by biexponential decay fitting. Right: Comparison of average relaxation rates between regions without and with Ub(Mn). The bar values are presented as mean relaxation rates $\pm$ standard deviation. Measured Ub(Mn)-induced acceleration $\Delta\Gamma$ is $(3.5\pm0.4)\times10^3\ \si{s}^{-1}$. %3551$\pm$371 s$^{-1}$. 
(B) Simulated relationship between relaxation acceleration and Ub(Mn) density. Black points: simulation data; blue line: linear fit. Gray shaded region: Ub density corresponding to the experimental $\Delta\Gamma=(3.5\pm0.4)\times10^3\ \si{s}^{-1}$. %3551$\pm$371 s$^{-1}$.
 }
\end{figure*}

\begin{figure*}[ht]
    \centering
\includegraphics[width=.85\textwidth]{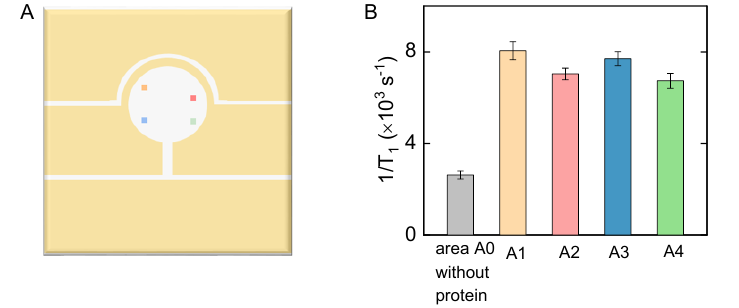}
\caption{\label{figs:PEIuniform} \textbf{Ub bonding uniformity on PEI-coated diamond surface.} (A) Overhead view of the coplanar waveguide around the diamond (yellow: gold plating; white: glass substrate). Four points represent locations of four regions used for uniformity test, separated by $\sim$500 $\upmu$m. (B) Spatial uniformity of relaxation rates. $T_1$ curves of 36 spots per region (marked in (A)) were measured to obtain the relaxation rates. Mean relaxation accelerations are $(5.4\pm0.4)\times10^3\ \si{s}^{-1}$, $(4.4\pm0.3)\times10^3\ \si{s}^{-1}$, $(5.1\pm0.4)\times10^3\ \si{s}^{-1}$, $(4.1\pm0.4)\times10^3\ \si{s}^{-1}$ relative to regions without proteins. The intra-regional and inter-regional fluctuations are 13$\%$ and 8$\%$, respectively. %$5427\pm 433\ \si{s}^{-1}$, $4413\pm 311\ \si{s}^{-1}$, $5076\pm 350\ \si{s}^{-1}$, $4113\pm368\ \si{s}^{-1}$. 
}
\end{figure*}

We quantitatively evaluated the density and uniformity of PEI functionalization through quantum relaxometry measurements. \autoref{figs:PEIdensity}A (left) shows the measured $T_1$ curves for diamond surfaces with and without Ub(Mn) bonding, revealing a mean relaxation rate acceleration of $(3.5\pm0.4)\times10^3\ \si{s}^{-1}$ (\autoref{figs:PEIdensity}A, right). 
Following the simulation described in the main text, we calculated the relaxation rate change ($\Delta\Gamma$) as a function of Ub(Mn) density (\autoref{figs:PEIdensity}B). The black dots are the simulated signal intensities $\Delta\Gamma$ at different bonding densities and the blue line is the linear fitting curve. The gray shaded area marks the Ub density corresponding to a signal intensity of $(3.5\pm0.4)\times10^3\ \si{s}^{-1}$, %3551$\pm$371 s$^{-1}$. 
yielding an average Ub(Mn) spacing of $9.3-10.4$ nm. 
To evaluate spatial uniformity, we performed $T_1$ measurements at 36 spots across four regions separated by $\sim500\ \upmu\si{m}$ (\autoref{figs:PEIuniform}A). After subtracting background relaxation rates, the mean accelerations were $(5.4\pm0.4)\times10^3\ \si{s}^{-1}$, $(4.4\pm0.3)\times10^3\ \si{s}^{-1}$, $(5.1\pm0.4)\times10^3\ \si{s}^{-1}$, $(4.1\pm0.4)\times10^3\ \si{s}^{-1}$ (\autoref{figs:PEIuniform}B). %$5427\pm 433\ \si{s}^{-1}$, $4413\pm 311\ \si{s}^{-1}$, $5076\pm 350\ \si{s}^{-1}$, $4113\pm368\ \si{s}^{-1}$. 
The observed intra-region variation ($8\%$) and inter-region deviation ($13\%$) indicate good spatial uniformity in protein bonding density across the functionalized surface.

\begin{figure*}[ht]
    \centering
\includegraphics[width=1.05\textwidth]{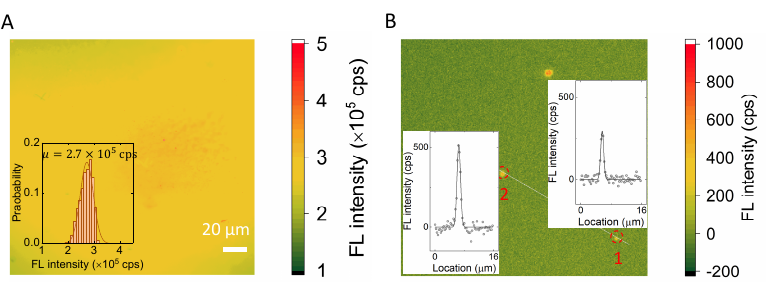}
\caption{\label{figs:PEIDensitySA488} {\textbf{Fluorescence characterization of PEI functionalized diamond surface.} (A) Fluorescence image of diamond surface coated with PEI and SA protein labeled with an Alexa Fluor 488 dye (SA-AF488). Inset: Histogram of pixels in the fluorescence image. The red line is the Gaussian fitting curve with $\mu=2.7\times10^5$ cps and $\sigma=0.4\times10^5$ cps. (B) Fluorescence of a single SA-AF488 protein. The right (left) inset shows $\sim$260 cps (520 cps) fluorescence intensity of the single (double) SA-AF488 marked in spot 1 (2). }
}
\end{figure*}

\begin{figure*}[ht]
    \centering
\includegraphics[width=.85\textwidth]{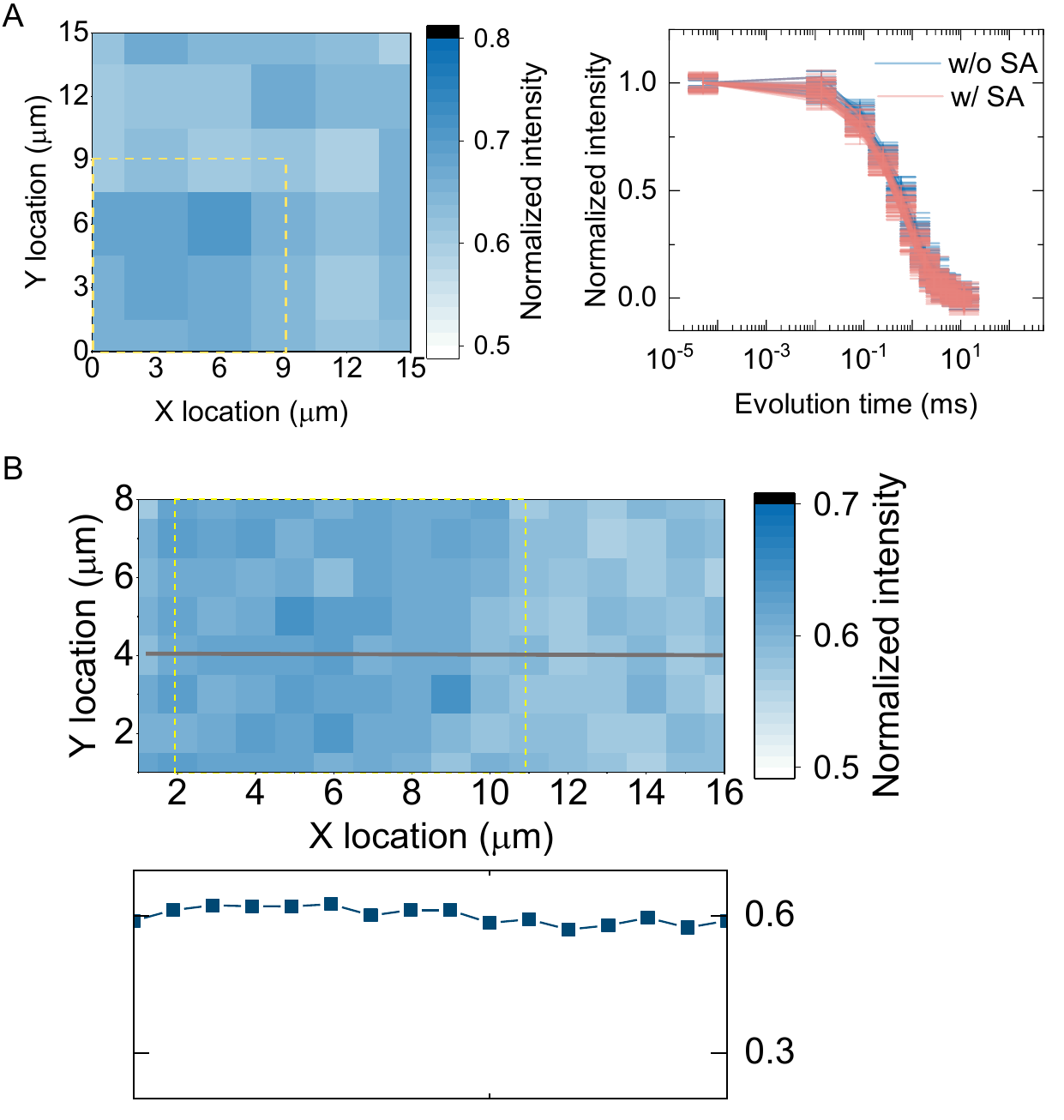}
\caption{\label{figs:SAsignal} \textbf{Relaxation results of diamond bonding with SA or BSA.} (A) Left: SA-bound surface fluorescence map measured by $T_1$ sequence with fixed $\tau=0.26\ \si{ms}$. Proteins in the region within the yellow dashed box were removed by AFM. Right: $T_1$ curves for regions with (pink) / without (blue) SA. (B) Top: BSA-bound surface fluorescence map measured by $T_1$ sequence with fixed $\tau=0.35\ \si{ms}$. Proteins in the region within the yellow dashed box were removed by AFM. Bottom: Fluorescence intensity profile along the gray line.
 }
\end{figure*}

We also used fluorescence labels to evaluated the protein bonding density. The diamond surface was bonded with PEI and SA protein labeled with an Alexa Fluor 488 dye (SA-AF488). The fluorescence imaging was detected with a wide-field inverted fluorescence microscope (Olympus, IRX5), in which the fluorescence was collected by an oil objective lens [Olympus MPLFLN 60$\times$] and captured by an EMCCD camera [Andor iXon Ultra897]. As shown in \autoref{figs:PEIDensitySA488}A, the fluorescence intensity of each pixel presented with peak$\pm$standard deviation is $(2.7\pm0.4)\times10^5$ cps. Besides, we dispersed the SA-AF488 protein solution and detected the fluorescence intensity of $\sim$260 cps of a single SA-AF488 (\autoref{figs:PEIDensitySA488}B). Considering the 16 $\upmu$m pixel size of the camera and the 60$\times$ magnification of the lens, the average density of SA-AF488 is $0.015\ \si{nm}^{-2}$ (i.e., spacing distance $\sim$8 nm), which is consistent with the result estimated using our relaxation method (12 nm).

\clearpage

\section{\label{sec:SAandBSA background} Background relaxation rate measurement of SA- and BSA-bound diamond}

\clearpage

\section{\label{sec:SPR} SPR analysis of SA/BSA interactions with biotin-Ub}
\begin{figure*}[ht]
    \centering
\includegraphics[width=.85\textwidth]{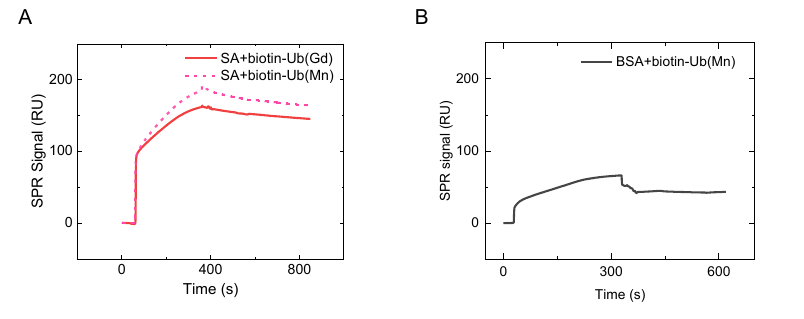}
\caption{\label{figs:spr} \textbf{SPR analysis of SA/BSA interactions with biotin-Ub.} (A) Binding kinetics of biotin-Ub(Gd) (solid line) and biotin-Ub(Mn) (dashed line) to SA protein. Endpoint signals for SA+biotin-Ub are 152 RU and 99 RU, the difference of which is attributed to SA density variations between sensor chip channels. (B) Binding kinetics of biotin-Ub(Mn) (solid line) to BSA protein. The endpoint signal is 43 RU.
 }
\end{figure*}

We employed surface plasmon resonance (SPR) on a Biacore$^{\text{TM}}$ T200 system with CM5 dextran-coated sensor chips to quantify interactions between SA or BSA and biotin-Ub complex. Before SA immobilization, CM5 chips were cleaned with the 4:1 mixture of 5$\%$ SDS and 50 mM NaOH, deionized water (ddH$_2$O), and dried with nitrogen gas N$_2(g)$. Carboxyl groups were activated via EDC/NHS treatment (15 min), followed by ethylenediamine conjugation (80 $\upmu\si{L}$, 20 min) to generate an amine-functionalized surface. Residual carboxyl groups were blocked with ethanolamine. NHS-PEG(2k)-biotin (4 mM, 80 $\upmu\si{L}$) was then reacted with the amine surface for 2 hours. Then the chip was installed inside the Biacore$^{\text{TM}}$ T200 platform. SA solution (30 $\upmu\si{g}/\si{mL}$, 100 $\upmu\si{L}$) was injected and bonded to biotin on the chip surface, thereby immobilizing the SA protein. In contrast, BSA proteins were directly immobilized on the activated carboxyl surface outside the Biacore$^{\text{TM}}$ T200 platform. Following the immobilization of SA or BSA, a 60 $\upmu$g/mL biotin-Ub solution was injected at 30 $\upmu$L/min to assess binding kinetics. SPR endpoint signals (152 RU/99 RU, \autoref{figs:spr}A) revealed strong SA+biotin-Ub interactions, compared to minimal BSA+biotin-Ub binding (43 RU, \autoref{figs:spr}B). Resonance unit (RU) values reflect angular shifts, where 1 RU corresponds to $1\ \si{RU}=10^{-4}\ ^\circ$.
\clearpage

\section{\label{sec:AFMBSAbioUb} AFM measurement of BSA+biotin-Ub coated diamond}
\begin{figure*}[htb]
    \centering
\includegraphics[width=0.95\textwidth]{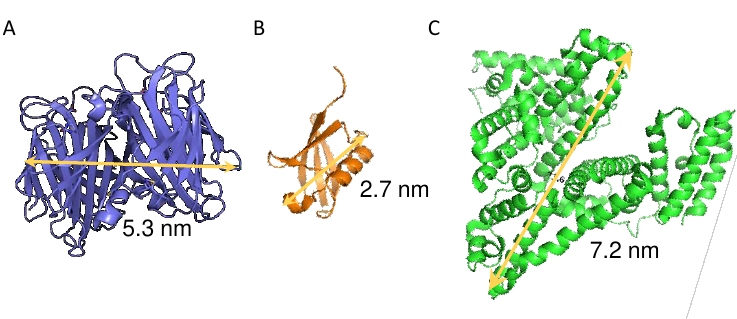}
\caption{\label{figs:protein3Dstructure} {\textbf{3D structures of (A) SA, (B) Ub and (C) BSA.} Figures are downloaded from protein data bank (SA: 1MEP, Ub: 1UBI, BSA: 4F5T). The yellow lines mark the average sizes of each protein.}}
\end{figure*}

\begin{figure*}[htb]
    \centering
\includegraphics[width=1\textwidth]{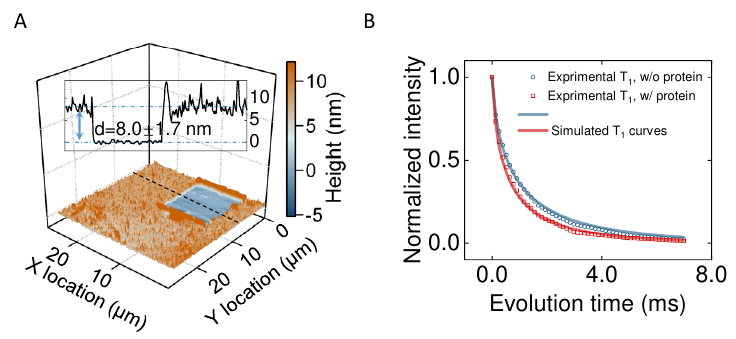}
\caption{\label{figs:BSAAFM} {(A) Topography image of the diamond surface coated with BSA and biotin-Ub(Mn) using AFM. Proteins in the square depression area were removed using AFM contact mode. The inset shows a total thickness of 8.0$\pm$1.7 nm. (B) Comparison of experimental and simulated $T_1$ curves when the diamond is coated with BSA+biotin-Ub(Mn). The solid red line is simulated at protein layer thickness $h = 8$ nm, and biotin-Ub density $\sigma=0.007\ \si{nm}^{-2}$.}}
\end{figure*}

\begin{figure}[hbt]
    \centering
\includegraphics[width=.9\textwidth]{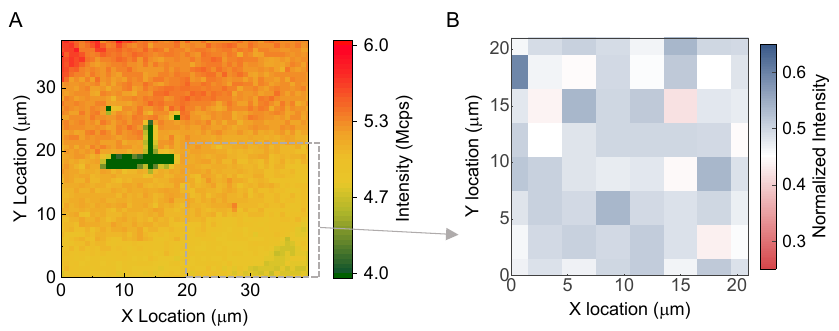}
\caption{\label{figs:136ensembleFLmap} \textbf{Fluorescence and relaxation rate distribution of ensemble NV centers.} (A) Fluorescence map under 532 nm laser excitation at 30 $\upmu$W. The dark cross area is an etched mark. (B) Relaxation rate map. The evolution time in $T_1$ measurement sequence is fixed at $\tau=350\ \upmu\si{s}$.}
\end{figure}

Similar to Figure 2B, We used AFM to measure the thickness of BSA+biotin-Ub layer coated on diamond surface (correlated relaxation data is shown in Figure 2F-H of main text) and the topography image indicates a thickness of $\sim$8 nm (\autoref{figs:BSAAFM}A) . Considering the average size of BSA is $\sim$7 nm (\autoref{figs:protein3Dstructure}C), we believed that the proteins on the diamond surface are monolayer. Though this monolayer is indeed higher than $\sim$6 nm of SA+biotin-Ub, the 2-nm thickness increase cannot explain the signal difference from $4.8\times10^3\ \si{s}^{-1}$ to $0.8\times10^3\ \si{s}^{-1}$. Furthermore, we simulated and found that the biotin-Ub density is about 0.007 nm$^{-2}$, assuming the BSA height of 8 nm (\autoref{figs:BSAAFM}B). This simulated biotin-Ub density ratio 0.028:0.007 (SA+biotin-Ub: BSA+biotin-Ub) is consistent with the SPR signal intensity ratio of $\sim$3:1.

\clearpage

\section{Diamond sample\label{sec:diamondsample}}

\subsection{Bulk diamond with ensemble NV centers\label{subsec:ensemblediamond}}

\subsection{Nano-pillar diamond with single NV centers\label{subsec:pillar446FLtest}}
\begin{figure*}[ht]
    \centering
\includegraphics[width=.9\textwidth]{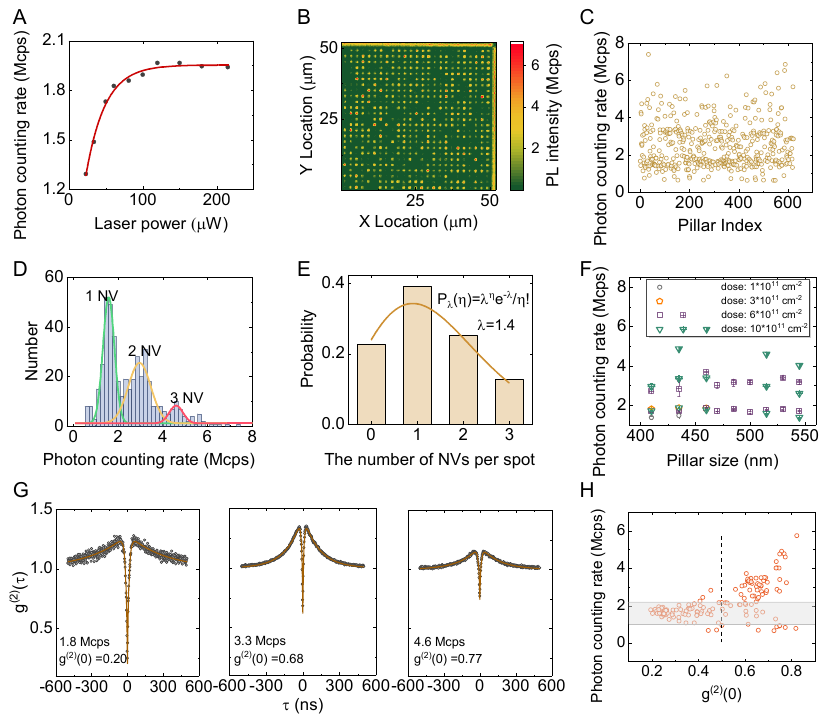}
\caption{\label{figs:446pillarFLmap} \textbf{Fluorescent properties of the nano-pillar diamond.} (A) Dependence of fluorescence counting rate of a single NV on 532 nm laser power (1.42 NA oil objective). (B) Fluorescent image of a pillar array (implantation dose: $1\times10^{12}\ \si{cm}^{-2}$) under 75 $\upmu\si{W}$ 532 nm excitation. (C) Background-subtracted NV count rates for pillars in (B). Pillars with counting rates below 500 kcps are considered to contain no NV centers and provide background fluorescence. (D) Histogram of counting rates in (C) with Gaussian fits (peaks: single [1.2 Mcps], double [2.4 Mcps], triple [3.5 Mcps] NV centers). (E) Probability distribution of NV number within a nano-pillar, derived from Figures (B) and (D). The solid yellow line is the fitting curve of the Poisson distribution. (F) Dependence of the average counting rate on the end size of nano-pillars. (G) Second-order correlation function $g^{(2)}(\tau)$ for three typical nano-pillars with distinct counting rates. (H) Relationship between $g^{(2)}(\tau)$ and fluorescence counting rate. The dotted line shows $g^{(2)}(\tau)=0.5$ and the gray shaded region marks the area with fluorescence counting rate between $1-2.2$ Mcps. }
\end{figure*}

The nano-pillar diamond used in this work was fabricated by a self-aligned patterning technique\cite{wmq2022sciadv}, via which NV centers are positioned within a 30 nm range from the center of each nano-pillar, resulting in a high fluorescence counting rate. The diamond was implanted with 4keV $^{15}\text{N}^+$ ions at four doses of $1\times10^{11}\ \si{cm}^{-2}$, $3\times10^{11}\ \si{cm}^{-2}$, $6\times10^{11}\ \si{cm}^{-2}$, $1\times10^{12}\ \si{cm}^{-2}$ (distributed across four zones). \autoref{figs:446pillarFLmap}A shows the dependence of the fluorescence counting rate of single NV centers in nano-pillars on the excited power of a 532 nm laser. The saturation counting rate is $\sim1.9\ $Mcps. \autoref{figs:446pillarFLmap}B displays the fluorescence image of an array with an implantation dose of $1\times10^{12}\ \si{cm}^{-2}$ under a laser power of 75 $\upmu\si{W}$. It can be seen that some nano-pillars exhibit low counting rates, indicating the absence of NV centers. Pillars with counting rates below 500 kcps are considered devoid of NV centers, and their average counting rate is defined as the background. \autoref{figs:446pillarFLmap}C presents the background-subtracted counting rates, which can be categorized into three intervals corresponding to the number of NV centers per pillar. \autoref{figs:446pillarFLmap}D is the statistical histogram of the data from \autoref{figs:446pillarFLmap}C. Three Gaussian peaks were obtained by fitting, and the peak positions of 1.5 Mcps, 3 Mcps, and 4.5 Mcps correspond to the average counting rates of single-NV, double-NV, and triple-NVs. This allows for an approximate inference of the number of NV centers in each nano-pillar based on the fluorescence counting rate. \autoref{figs:446pillarFLmap}E divides the counting rate intervals according to the Gaussian fitting results from \autoref{figs:446pillarFLmap}D, and the pillars without NV center are also taken into account. By fitting the probability distribution of NV center counts in 625 nano-pillars from \autoref{figs:446pillarFLmap}B with the Poisson function, an average number of NV centers $\lambda=1.4$ is obtained. Similarly, a similar analysis was performed on 19 arrays with different four implantation doses and nine end sizes of nano-pillars, yielding the average counting rates for single- (double-, triple-) NV centers (\autoref{figs:446pillarFLmap}F). Overall, the nano-pillar size (400-500 nm) has a negligible impact on the average counting rate. And minor variations likely arise from deviations in ion implantation alignment. In \autoref{figs:446pillarFLmap}F, implantation doses are color-coded, showing that the dose also has minimal influence on the average counting rate. \autoref{figs:446pillarFLmap}G shows the second-order correlation function $g^{(2)}(\tau)$ measured for nano-pillars with three typical counting rates. The counting rates align with the general criterion $g^{(2)}(\tau)<0.5$ for single NV centers. \autoref{figs:446pillarFLmap}H shows $g^{(2)}(\tau)$ against the fluorescence counting rate for 187 NV centers. The dashed line at $g^{(2)}(\tau)=0.5$ serves as the criterion for distinguishing single from multiple NV centers. The gray shaded region marks the fluorescence counting rate range of 1-2.2 Mcps for single NV centers based on counting rate analysis. Of the 83 pillars within this range, 72 exhibit $g^{(2)}(\tau)<0.5$, yielding an accuracy rate of 87$\%$ for identifying single NV centers using the counting rate as the criterion. Consequently, pillars with count rates between 1–2.2 Mcps were selected as sensors for further experiments.

\section{\label{sec：T1rate} Theoretical derivation of the relaxation rate}
In this section, we will derive the NV relaxation and magnetic noise signal strength in detail. Some results are derived from references, which are restated here briefly.
\subsection{\label{subsec：T1crv}$T_1$ curve}
The polarized NV spin state relaxes gradually to the thermal equilibrium state under the influence of ambient noise. The relaxation process includes single and double quantum processes, considering the effect of magnetic and electrical noise \cite{DQT1PRL2017}. At room temperature, the transition rate between $m_S=0$ and $m_S=\pm1$ is defined as $k_{01}$, and the transition rate between $m_S=+1$ and $m_S=-1$  is defined as $k_{11}$, then the change of the population of each spin state $n_{0,\pm1}(t)$  are determined by the rate equation\cite{DQT1PRL2017,Gd2013PRB}
\begin{equation}
    \frac{d}{d t}\left(\begin{array}{c}
n_0 \\
n_{-1} \\
n_{+1}
\end{array}\right)=\left(\begin{array}{ccc}
-2 k_{01} & k_{01} & k_{01} \\
k_{01} & -k_{01}-k_{11} & k_{11} \\
k_{01} & k_{11} & -k_{01}-k_{11}
\end{array}\right)\left(\begin{array}{c}
n_0 \\
n_{-1} \\
n_{+1}
\end{array}\right).
\end{equation}
Combining the initial conditions $n_0|_{t=0}=n_0(0)$, $ n_{-1}|_{t=0}=n_{-1}(0)$, $n_{+1}|_{t=0}=n_{+1}(0)$, the equal occupation probability at thermal equilibrium $n_0|_{t=\infty}=n_{-1}|_{t=\infty}=n_{+1}|_{t=\infty}=1/3$ and the 
constraint $n_0(t)+n_{-1}(t)+n_{+1}(t)=1$, the solution of the above rate equation is
\begin{equation}
    \begin{dcases}
    n_{0}(t)=\frac{1}{3}+\left(n_{0}(0)-\frac{1}{3}\right)e^{-3k_{01}t},\\
    n_{\pm1}(t)=\frac{1}{3}\pm\frac{n_{+1}(0)-n_{-1}(0)}{2}e^{-(k_{01}+2k_{11})t}-\frac{1}{2}\left(n_{0}(0)-\frac{1}{3}\right)e^{-3k_{01}t}.\\    
    \end{dcases}
\end{equation}
Therefore, the normalized $T_1$ curve obtained with the sequence in the main text (Figure 1B) is $I(t)=\left[\frac{1}{3}+(1-\frac{1}{3})e^{-3k_{01}t}\right]-\left[\frac{1}{3}+(0-\frac{1}{3})e^{-3k_{01}t}\right]=e^{-3k_{01}t}$. Therefore,
\begin{equation}
    1/T_1=3k_{01}.
\end{equation}

\subsection{\label{subsec:T1rate}Longitudinal relaxation rate $1/T_1$ }
The fast spin-flips of electron spins with short $T_1$ at room temperature provide a non-zero RMS fluctuating magnetic field at the location of the NV center, thus accelerating the dipole transition between the spin sub-levels. So the magnetic noise signal of the metal-based spin labels can be detected by measuring the longitudinal relaxation rate of the NV center. Specifically,
\begin{equation}
    \Gamma_1=\frac{1}{T_1}=\Gamma_{1,\text{BG}}+\Gamma_{1,\text{induced}},
\end{equation}
where $\Gamma_{1,\text{BG}}$ is the background relaxation rate, which is affected by lattice relaxation, defect density near the NV center, and other factors like temperature and magnetic field. $\Gamma_{1,\text{induced}}$ is contributed by metal spin labels. For NV center ($S=1$), the transition rate\cite{Gd2013PRB,principle1990slichter}
\begin{equation}
    k_{01}=\frac{1}{2}\gamma_\text{NV}^2 S_{B_\bot}(\omega_0),
\end{equation}
where $\gamma_\text{NV}$ is the gyromagnetic ratio of NV center and $S_{B_\bot}(\omega)$ is the spectral density function of the transverse magnetic noise. \\
\\
\textbf{Spectral density function} 

The autocorrelation function is $\langle B_\bot(t)B_\bot(t+\tau)\rangle = \langle B_\bot^2 \rangle \exp\left(-|\tau/\tau_c|\right)\cos\left(\omega_L\tau\right)$. According to the Weiner-Khinchin theorem, a spectral density function can be obtained by a Fourier transform of the autocorrelation function\cite{2020PRBspecdens}, thus
\begin{equation}
\begin{aligned}
     S_{B_\bot}(\omega) &= \int \text{d}\tau \langle B_\bot^2 \rangle \exp\left(-|\tau/\tau_c|\right)\cos\left(\omega_L\tau\right) e^{-\text{i}\omega\tau} \\
    &=\langle B_\bot^2\rangle \left[\frac{\tau_c}{1+(\omega-\omega_L)^2\tau_c^2}+\frac{\tau_c}{1+(\omega+\omega_L)^2\tau_c^2}\right].
\end{aligned}    
\end{equation}
At zero magnetic field, when the broadening is much larger than the spin label's zero-field splitting, the spectral density function can be approximated as
\begin{equation}
    S_{B_\bot}(\omega) \approx  \frac{2\tau_c \langle B_\bot^2\rangle}{1+\left(\omega_0\tau_c\right)^2},
\end{equation}
where $\omega_0/2\pi=2.87\ $GHz is the zero-field splitting of the NV center.\\
\\
\textbf{Transverse magnetic noise} 

According to the dipole-dipole interaction, the magnetic field $\bm{B}_i$  from a nearby spin $\bm{S}_i$ detected by NV center at the position $\bm{r}_i$ is 
\begin{equation}
   \bm{B}_i=\frac{\mu_0\hbar\gamma_i}{4\pi r_i^3}\left[\bm{S}_i-3\left(\bm{S}_i\cdot\frac{\bm{r}_i}r_i{}\right)\frac{\bm{r}_i}{r_i}\right],
\end{equation}
where $\gamma_i$ is the gyromagnetic ratio of the spin label. Then the magnetic field operators are given by
\begin{equation}
    \mathcal{B}_{i,x}=\frac{\mu_0\hbar\gamma_i}{4\pi r_i^3}\left[\mathcal{S}_{i,x}-3\sin{\theta_i}\cos{\phi_i}\left(\bm{S}_i\cdot \bm{r}_i\right)\right],
\end{equation}
\begin{equation}
    \mathcal{B}_{i,y}=\frac{\mu_0\hbar\gamma_i}{4\pi r_i^3}\left[\mathcal{S}_{i,y}-3\sin{\theta_i}\sin{\phi_i}\left(\bm{S}_i\cdot \bm{r}_i\right)\right],
\end{equation}
\begin{equation}
    \mathcal{B}_{i,z}=\frac{\mu_0\hbar\gamma_i}{4\pi r_i^3}\left[\mathcal{S}_{i,z}-3\cos{\theta_i}\left(\bm{S}_i\cdot \bm{r}_i\right)\right],
\end{equation}
where $\bm{r}_i = r_i(\sin{\theta_i}\cos{\phi_i},\sin{\theta_i}\sin{\phi_i},\cos{\theta_i})$, $\theta_i$ and $\phi_i$ are the parameters of the spherical coordinate frame. Therefore, the transverse magnetic field can be evaluated as
\begin{equation}
    \langle B_{i,\bot}^2\rangle =Tr\left\{\rho_\text{th}\left(\mathcal{B}_{i,x}^2+\mathcal{B}_{i,y}^2\right)\right\}= \frac{S(S+1)}{3}\left(\frac{\mu_0\gamma_i\hbar}{4\pi}\right)^2\frac{2+3\sin^2{\theta_i}}{r_i^6},
\end{equation}
where $S$ is the spin quantum number of labels, $\rho_\text{th}$ is the density matrix of thermal equilibrium state. The magnetic field intensity exhibits exceptional sensitivity to the separation distance between the NV center and the spin label, rendering this approach particularly effective for achieving high-resolution magnetic imaging. Leveraging this principle, researchers have successfully demonstrated magnetic scanning with a remarkable 10-nanometer spatial resolution
\cite{wpf2019sciadFe}.

In summary, the acceleration of NV relaxation by the spin bath is 
\begin{equation}
\label{eqs:gammainduced}
\begin{aligned}
    \Gamma_{1,\text{induced}}&=\sum_i \Gamma_{1,i} \\
    &=\sum_i \frac{3}{2}\frac{S(S+1)}{3}\left(\frac{\mu_0\gamma_i\gamma_\text{NV}\hbar}{4\pi}\right)^2\frac{2+3\sin^2{\theta_i}}{r_i^6} \frac{2\tau_c}{1+\left(\omega_0\tau_c\right)^2} \\
    &=\sum_i \frac{3b^2_{i,\bot}\tau_c}{1+\left(\omega_0\tau_c\right)^2},
    \end{aligned}
\end{equation}
where
\begin{equation}
\label{eqs:b2coupling}
    b_{i,\bot}^2=\frac{S(S+1)}{3}\left(\frac{\mu_0\gamma_i\gamma_\text{NV}\hbar}{4\pi}\right)^2\frac{2+3\sin^2{\theta_i}}{r_i^6}
\end{equation}
is used to describe the coupling strength between the NV center and the spin label.

\clearpage

\section{\label{sec:ensembleNVT1model} $T_1$ curve simulation of ensemble shallow NV center}
\subsection{$T_1$ curve of ensemble NV subtracting charge state relaxation \label{subsec:T1crvexcludeCharge}}
\begin{figure*}[ht]
    \centering
\includegraphics[width=.9\textwidth]{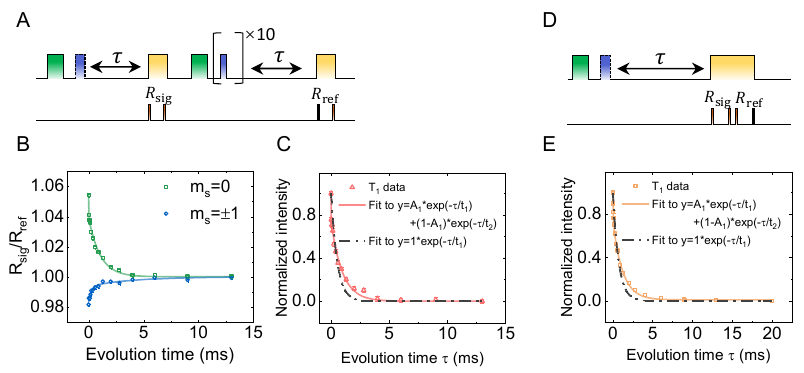}
\caption{\label{figs:t1crvchargedelete} \textbf{$T_1$ curve of ensemble NV centers with charge state relaxation subtraction.} (A) $T_1$ measurement sequence designed to eliminate the influence of charge state decay with thermal equilibrium spin state preparation. Green (yellow) pulses are 532 (594) nm laser, and blue pulses denote microwave pulses. The lower red pulses are triggers for data acquisition (DAQ). $R_\text{sig}$ captures both charge and spin relaxation processes, whereas $R_\text{ref}$ contains only charge state decay information. (B) Relaxation curves with initial spin state $m_S=0$ (green dots) or $\pm1$ (blue dots). Solid curves are fitting curves with biexponential functions. (C) Normalized $T_1$ relaxation curve. Data points are derived from the difference between the two curves in (B) and normalized to the first $\tau$ point. The dashed (solid) line is the fitting curve with (bi-)exponential decay function. (D) $T_1$ measurement sequence eliminating influence of charge state decay using 594 nm laser to readout the charge state. (E) Normalized $T_1$ curve. The dashed (solid) line is the fitting curve with (bi-)exponential decay function.  
}
\end{figure*}
As shown in Figure 5A of the main text, ensemble NV centers exhibit characteristic biexponential $T_1$ decay, consistent with previous reports\cite{sciadv2021radical,acscent2023radical,redox2022radical,nanolett2013Mn}. The fast decay component is usually attributed to cross-relaxation of NV pairs or influence of NV centers close to diamond surface. In this work, we exclude the former possibility due to the low nitrogen implantation dose in the diamond (dose $1\times 10^{13}\ \si{cm}^{-2}$, corresponding to an estimated average NV separation of $\sim 30$ nm). The contribution of the latter will be discussed below. Here, we focused on isolating the influence of charge state relaxation to the $T_1$ process. A 594 nm laser, which minimally excites fluorescence from the $\text{NV}^0$, was employed to read out the NV spin state during $T_1$ measurement with ensemble NV centers. As shown in \autoref{figs:t1crvchargedelete}A, NV center was initialized into the spin state $m_S=0$ ($m_S=\pm1$) by a 532 nm laser ( and the following microwave $\pi-$pulse). After a waiting time $\tau$, a 594 nm laser was used to read out the fluorescence $R_\text{sig}$, associating with spin population. Assuming fluorescence counts for $m_S=0$ and $m_S=\pm1$ are $I_0$ and $I_{\pm1}$, and the population of charge state $\text{NV}^-$, spin state $m_S=0$ and spin state $m_S=\pm 1$ are $p_{\text{NV}^-}(\tau)$, $p_{m_S=0}(\tau)$, $p_{m_S=\pm 1}(\tau)$ at time $\tau $, fluorescence intensity can be expressed as $R_\text{sig}=p_{\text{NV}^-}(\tau)\cdot \left(p_{m_S=0}(\tau)I_0+p_{m_S=\pm 1}(\tau)I_{\pm1}\right)$. In the second half of the sequence, the thermal equilibrium state $\rho_\text{th}$ was prepared using repeated ‘0.61$\pi$-waiting' pulses. After the same waiting time $\tau$, fluorescence intensity $R_\text{ref}=p_{\text{NV}^-}(\tau)\cdot \left(1/3I_0+2/3I_{\pm1}\right)$ was excited by 594 nm laser as the reference. Two curves of the ratio $R_\text{sig}/R_\text{ref}$ were obtained, with charge state relaxation term $p_{\text{NV}^-}$ effectively removed (\autoref{figs:t1crvchargedelete}B). This allows extraction of normalized charge-independent $T_1$ relaxation curve (\autoref{figs:t1crvchargedelete}C). Similarly, \autoref{figs:t1crvchargedelete}D shows an alternative sequence to eliminate charge state influence using a 594 nm laser for reading out and the $T_1$ curve is presented in \autoref{figs:t1crvchargedelete}E. Both $T_1$ curves (\autoref{figs:t1crvchargedelete}C, E) consistently require biexponential fitting, with a persistent fast decay component. This observation suggests the biexponential decay shape of ensemble NV $T_1$ is not attributed to simple charge fluctuations.

\subsection{\label{subsec:surfacemagnoise}Surface magnetic noise model of shallow NV centers}
\begin{figure*}[ht]
    \centering
\includegraphics[width=.9\textwidth]{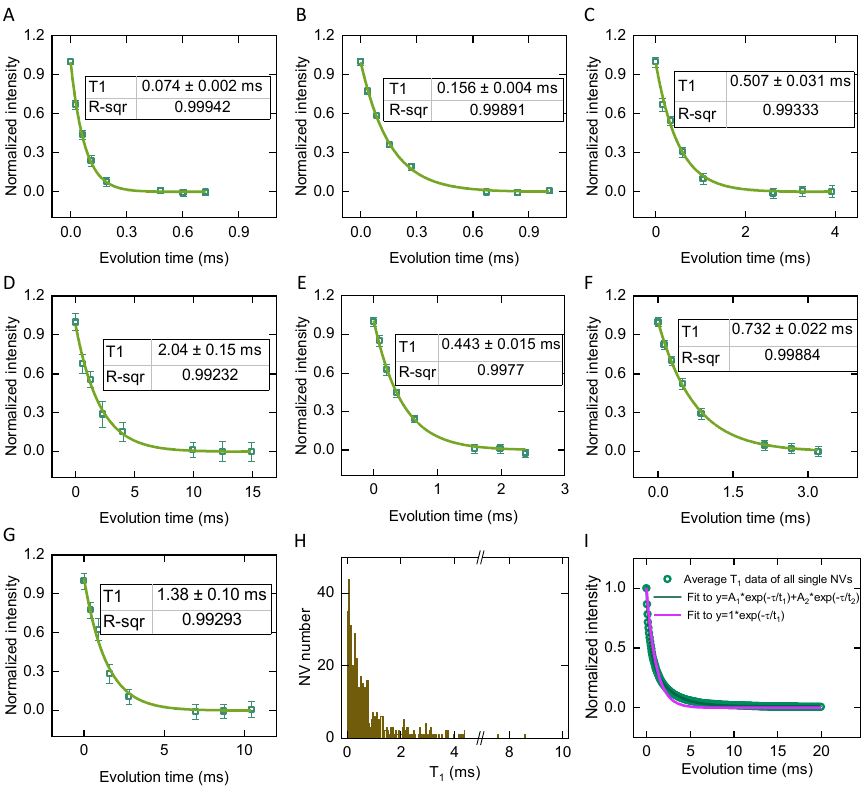}
\caption{\label{figs:singleNVt1crvs} \textbf{$T_1$ curves of single NV centers.} (A-G) 7 typical $T_1$ curves of single NV centers. Green dots are experimental data points, and solid lines are fitting curves with a single exponential decay function. (H) Histogram of $T_1$ data for 471 single NV centers. (I) Averaged $T_1$ curves of 471 single NV centers in (H) (green dots). The pink (Green) line is a fitting curve to (bi-)exponential decay function.
}
\end{figure*}

Generally, $T_1$ curves of single NV centers fit good to the single exponential decay function (\autoref{figs:singleNVt1crvs}A-G). Statistical analysis of $T_1$ data fitting from 471 single NV centers (\autoref{figs:singleNVt1crvs}H) reveals a broad distribution of relaxation times. The averaged $T_1$ curve of these NV centers fits better to the biexponential decay function (\autoref{figs:singleNVt1crvs}I), suggesting that the heterogeneity in single NV relaxation rates collectively contributes to the biexponential shape observed at the ensemble level.

\begin{figure*}[hbt]
    \centering
\includegraphics[width=.9\textwidth]{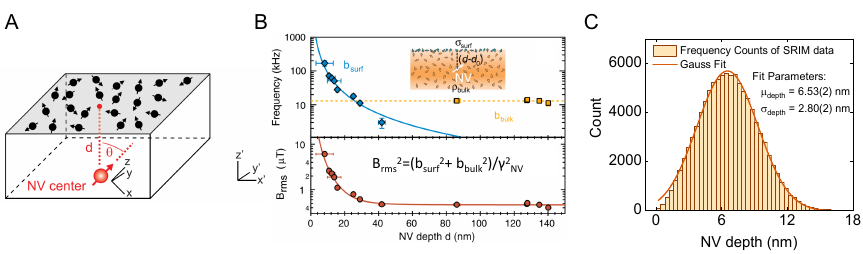}
\caption{\label{figs:surfacenoisemodel} \textbf{The model of magnetic noise for shallow NV centers.} (A) Schematic of paramagnetic defects on diamond surface\cite{PRL2014surfacenoise}. (B) 
 Upper panel: depth dependence of NVs' coupling to the surface magnetic impurities and bulk noise. $b_\text{surf}\propto d^{-2}$, while $b_\text{bulk}$ is independent with depth. Lower panel: total RMS magnetic field of two spin bath\cite{PRL2014myerssurface}. (C) Histogram of NV depth simulated with SRIM. The implantation energy is 4 keV $^{15}\text{N}$. The solid line is the Gaussian fitting curve.
 }
\end{figure*}

Furthermore, we investigated reasons for the diversity in relaxation rates among shallow single NV centers. The paramagnetic impurities on the diamond surface constitute a significant magnetic noise source that influences the relaxation process\cite{PRL2014surfacenoise,PRL2014myerssurface}. \autoref{figs:surfacenoisemodel}A illustrates a schematic of these relatively evenly distributed magnetic impurities on the diamond surface. The auto-correlation time of the noise spectrum is $\tau_{c,\text{surf}}=10^{-11}-10^{-5}\ $s\cite{PRL2014surfacenoise}. Using [$x', y', z'$] as the coordinate system, the distance between the impurity at [$x', y', d$] and the NV center is denoted as $r'$, while the angle between the impurity and the quantum axis of NV is $\theta'$. The angle between the principal axis of NV and the normal direction of the diamond surface is $\theta$, which can be $54.7^\circ$ or $0^\circ$, depending on the diamond cutting. According to the \autoref{eqs:gammainduced} and \autoref{eqs:b2coupling}, the total coupling strength for a shallow single NV center with depth of $d$ [nm] and a surface impurity density $\sigma_\text{surf}$ [spins$\cdot$nm$^{-2}$] is given by
\begin{equation}
\begin{aligned}
    b_{\text{surf},\bot}^2 &= \frac{S(S+1)}{3}\left(\frac{\mu_0\gamma_i\gamma_\text{NV}\hbar}{4\pi}\right)^2 \int_{-\infty}^{+\infty}\text{d}x'\int_{-\infty}^{+\infty}\text{d}y'\sigma_{\text{surf}}\cdot\frac{2+3\sin^2{\theta'}}{r'^6} \\
    &=
    \begin{dcases}
        \frac{2\pi A^2\sigma_\text{surf}}{d^4},\ \theta=54.7^\circ \\
        \frac{3\pi A^2\sigma_\text{surf}}{2d^4},\ \theta=0^\circ ,
    \end{dcases}
\end{aligned}
\end{equation}
where $A^2=\frac{S(S+1)}{3}\left(\frac{\mu_0\gamma_i\gamma_\text{NV}\hbar}{4\pi}\right)^2$. Therefore, the total magnetic field generated by the surface impurities is inversely proportional to the depth of the NV center. In addition, shallow NV center relaxation is also affected by bulk noise bath, whose effect is independent of NV center's depth. So the background relaxation rate ($\theta=54.7^\circ$) can be written as 
\begin{equation}
\label{eqs:surfT1}
\begin{aligned}
     \Gamma_{1,\text{BG}}=\frac{1}{T_{1,\text{BG}}} &=\Gamma_{1,\text{bulk}}+\Gamma_{1,\text{surf}} \\
     &=\Gamma_{1,\text{bulk}}+\frac{3\tau_{c,\text{surf}}}{1+\left(\omega_0\tau_{c,\text{surf}}\right)^2}\frac{2\pi A^2\sigma_\text{surf}}{d^4}.
\end{aligned}   
\end{equation}
\autoref{figs:surfacenoisemodel}B presents results from Ref.\citenum{PRL2014myerssurface}, where solid lines are fitting curves based on the model described above. On the other hand, the depth of single NV centers exhibits a wide broadening, as simulated by SRIM ( \autoref{figs:surfacenoisemodel}C). The fitting solid line to a Gaussian function indicates a center depth of 6.5 nm and a half-height full width of 6.6 nm. In conclusion, we attributed the broadening of relaxation rates in shallow NV centers to the variation in their depths, thus resulting in the biexponential decay shape of the $T_1$ curve of shallow ensemble NV center.
\subsection{\label{subsec:T1simuensemble}$T_1$ curve simulation of the ensemble shallow NV centers}

\begin{figure*}[bht]
    \centering
\includegraphics[width=.9\textwidth]{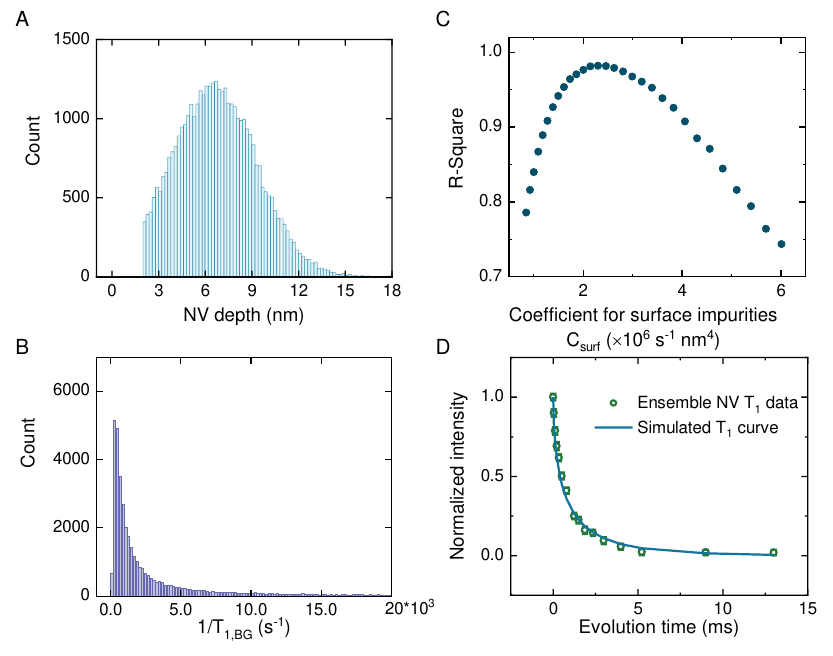}
\caption{\label{figs:136T1simulation} \textbf{$T_1$ curve simulation of ensemble shallow NV centers.} (A) Depth distribution of the simulation. (B) 
 A typical relaxation rate distribution according to the surface magnetic noise model at a surface impurity density of $\sigma_\text{surf}=0.40\ \si{nm}^{-2}$. (C) Dependence of goodness-of-fit (R-squared) on the coefficient $C_\text{surf}$, which is proportional to the surface impurity density $\sigma_{\text{surf}}$. The maximum R-squared value corresponds to the optimal $\sigma_\text{surf}=0.40\ \si{nm}^{-2}$. (D) Comparison of experimental $T_1$ data (green dots) with the simulated $T_1$ curve at $\sigma_\text{surf}=0.40\ \si{nm}^{-2}$. 
 }
\end{figure*}

We simulated $T_1$ curves for ensemble NV centers based on the surface magnetic noise model described above. The depth of NV centers were modeled as a normal distribution (peak location $\mu=6.5\ $nm, variance $\sigma=2.8\ $nm, and truncated depth > 2 nm considering shallow NVs' instability\cite{PRB2012shallowNVdegen}), as obtained by the SRIM simulation (\autoref{figs:136T1simulation}A). According to \autoref{eqs:surfT1}, the background relaxation rate of a shallow single NV center can be simplified as
\begin{equation}
\label{eqs:T1BG}
    \Gamma_{1,\text{BG}}=\Gamma_{1,\text{bulk}}+\frac{C_\text{surf}}{d^4},
\end{equation}
 where $C_\text{surf}\propto\sigma_\text{surf}$, and $\sigma_\text{surf}$ is the mean density of surface paramagnetic impurities. Considering the $T_1$ distribution of single NVs in \autoref{figs:singleNVt1crvs}H and Ref.\citenum{PRL2014surfacenoise}, we set $\Gamma_{1,\text{bulk}}=100\ \si{s}^{-1}$. Then relaxation rates of all NVs were calculated, with a typical distribution presented in \autoref{figs:136T1simulation}B. Simulated ensemble $T_{1}$ curves were generated by averaging single-exponential decays of individual NVs. The optimal surface impurity density ($\sigma_{\text{surf}}=0.40\ \si{nm}^{-2}$) was determined by maximizing the R-squared value between simulated and experimental curves (\autoref{figs:136T1simulation}C), assuming a surface noise autocorrelation time $\tau_{c,\text{surf}}=0.28\ \si{ns}$\cite{PRL2014surfacenoise}. As shown in \autoref{figs:136T1simulation}D, the simulated $T_1$ curve (solid line) closely matches experimental data (dots), with minor deviation arising from approximations in the depth distribution. Overall, this noise model effectively simulates $T_1$ curves of shallow ensemble NV centers.

%\clearpage
\section{\label{sec:pillarHistogmSimu}Simulation of signals detected by single NV centers}
\subsection{\label{subsec:pillarSurfimpurityDense}Density of surface paramagnetic impurities}

\begin{figure*}[htb]
    \centering
\includegraphics[width=.9\textwidth]{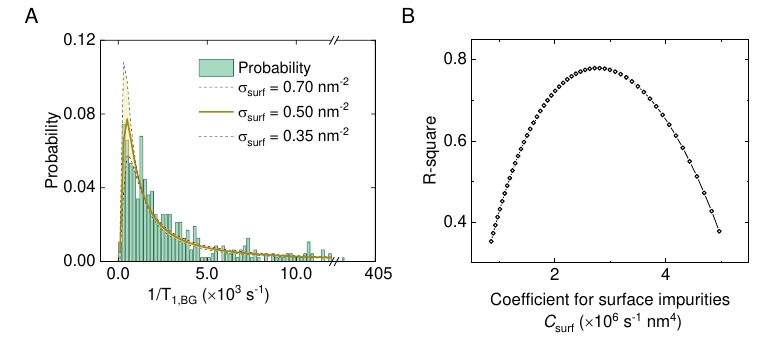}
\caption{\label{figs:446RateHistgmbare} \textbf{Analysis of background relaxation rates of shallow single NV centers.} (A) Background relaxation rate distribution for 471 shallow single NV centers (bare diamond). Simulated curves (yellow/green/red) correspond to surface defect densities $\sigma_\text{surf}=0.50\ \si{nm}^{-2}$ (R$^2$=0.78), $0.35\ \si{nm}^{-2}$ and $0.70\ \si{nm}^{-2}$ (both R$^2$=0.7) for magnetic noise autocorrelation time $\tau_{c,\text{surf}}=0.28\ \si{ns}$\cite{PRL2014surfacenoise}, assuming Gaussian NV depth ($\mu=5.5\ $nm, $\sigma=2.2\ $nm). (B) Dependence of the goodness-of-fit (R$^2$) on surface noise coefficient $C_\text{surf}$. Optimal agreement (R$^2$ =0.78) occurs at $C_\text{surf}=2.7\times10^6\ \si{s}^{-1}\cdot\si{nm}^4$, corresponding to $\sigma_\text{surf}=0.50\ \si{nm}^{-2}$.}
\end{figure*}

According to the model in \autoref{subsec:surfacemagnoise}, the background relaxation rates of shallow single NV centers exhibit a distribution due to the broadening of depth and the presence of relatively uniform surface magnetic noise. The histogram in \autoref{figs:446RateHistgmbare}A shows the probability distribution of background relaxation rates for 471 single NV centers in nano-pillar diamond. Based on the SRIM simulation for 4 keV $^{15}\text{N}^+$ implantation, depth distribution is approximately Gaussian with a peak at 6.5 nm and a full width at half maximum of 6.6 nm. Post-annealing (580 $^\circ$C in air, 20 min) shifted the depth distribution to $\mu=5.5\ $nm, $\sigma=2.8\ $nm. Combining \autoref{eqs:T1BG} with depth distribution, the probability distribution function of $\Gamma_{1,\text{BG}}$ is given by
\begin{equation}  f(\Gamma_{1,\text{BG}};\mu,\sigma,C_\text{surf},\Gamma_{1,\text{bulk}})=\frac{C_\text{nrm}}{4\sigma \sqrt{2\pi}}\frac{{C_\text{surf}}^{1/4}}{\left(\Gamma_{1,\text{BG}}-\Gamma_{1,\text{bulk}}\right)^{5/4}} \exp\left[\frac{\left(\left(\frac{C_\text{surf}}{\Gamma_{1,\text{BG}}-\Gamma_{1,\text{bulk}}}\right)^{1/4}-\mu\right)^2}{2\sigma^2}\right],
\end{equation}
with $C_\text{nrm}$ as a normalization constant introduced by histogram binning. This allows the relaxation rate distribution to be calculated for different values of $C_\text{surf}$ ($\propto\sigma_\text{surf}$). Comparing the goodness-of-fit between simulated and experimental distributions, the optimal value of $C_\text{surf}$ was determined to be $2.7\times10^6\ \si{s}^{-1}\cdot\si{nm}^4$ (\autoref{figs:446RateHistgmbare}B), corresponding to $\sigma_\text{surf}\sim0.50\ \si{nm}^{-2}$ for $\tau_{c,\text{surf}}=0.28\ \si{ns}$\cite{PRL2014surfacenoise}. The solid yellow line in \autoref{figs:446RateHistgmbare}A is the best-fit simulated distribution (R-squared=0.78), while simulations for $\sigma_\text{surf}=0.35\ \si{nm}^{-2}$ (red) and $0.70\ \si{nm}^{-2}$ (green) show a little reduced agreement (R-squared=0.7).

\subsection{\label{subsec:pillarSAbondDense}Streptavidin Bonding Density detected by single NVs in nano-pillars}
\begin{figure*}[htb]
    \centering
\includegraphics[width=.9\textwidth]{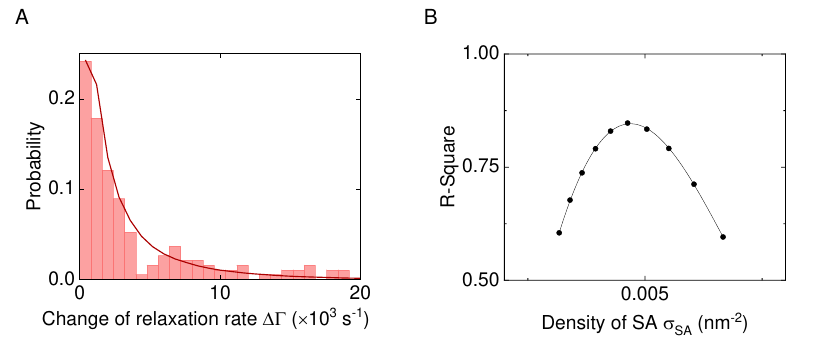}
\caption{\label{figs:446RateHistgmSAbioUB} \textbf{Analysis of signal for SA+biotin-Ub(Mn) with single NV centers.} (A) Probability distribution of relaxation rate changes for 195 single NV centers when nano-pillar diamond is coated with SA+biotin-Ub(Mn). The solid line is the simulated curve at SA bonding density $\sigma_\text{SA}=0.007\ \si{nm}^{-2}$. (B) Dependence of goodness-of-fit on SA bonding density. The black line is a cubic function fitting curve.}
\end{figure*}

We employed 195 single shallow NV centers (background relaxation rate $\Gamma_{1,\text{BG}}$<2000 s$^{-1}$) to detect the interaction between SA and biotin-Ub(Mn) complex. Probability distribution of signal intensity $\Delta\Gamma$ is shown in \autoref{figs:446RateHistgmSAbioUB}A. Data with signal intensity below 10,000 s$^{-1}$ were used to analyze the agreement between simulated and experimental distributions, as larger signals likely originating from local protein aggregation. The solid line in \autoref{figs:446RateHistgmSAbioUB}A represents the best-fit Monte Carlo simulation. The simulation was conducted using 10,000 shallow single NV centers with equal 100 nm lateral spacing and Gaussian depth distribution ($\mu=5.5\ $nm, $\sigma=2.8\ $nm), under a surface magnetic impurity density of $\sigma_\text{surf}=0.50\ \si{nm}^{-2}$. SA proteins were approximated as 5.8 nm cubes capable of binding four biotin-Ub(Mn) complexes, which were simplified as points with four Mn(II) labels exhibiting 3 GHz spectral broadening. Binding sites were positioned at opposing vertices on the top/bottom cube faces. Protein interactions were computed within $40\ \si{nm}\times40\ \si{nm}$ regions around each NV, assuming an average SA surface density ($\sigma_\text{SA}$). Relaxation rate changes were analyzed for NVs with $\Gamma_{1,\text{BG}}$<2000 s$^{-1}$, enabling statistical derivation of signal intensity distributions. \autoref{figs:446RateHistgmSAbioUB}B shows the goodness-of-fit between the simulated and experimental distributions, yielding a best-fit SA bonding density of $\sim0.007\ \si{nm}^{-2}$, which means the average protein spacing of $\sim12\ $nm.
%\clearpage

\section{Relaxation measurement under solution conditions}
\subsection{Molecular interaction detection under solution conditions}
\begin{figure}[htb]
    \centering
\includegraphics[width=1\textwidth]{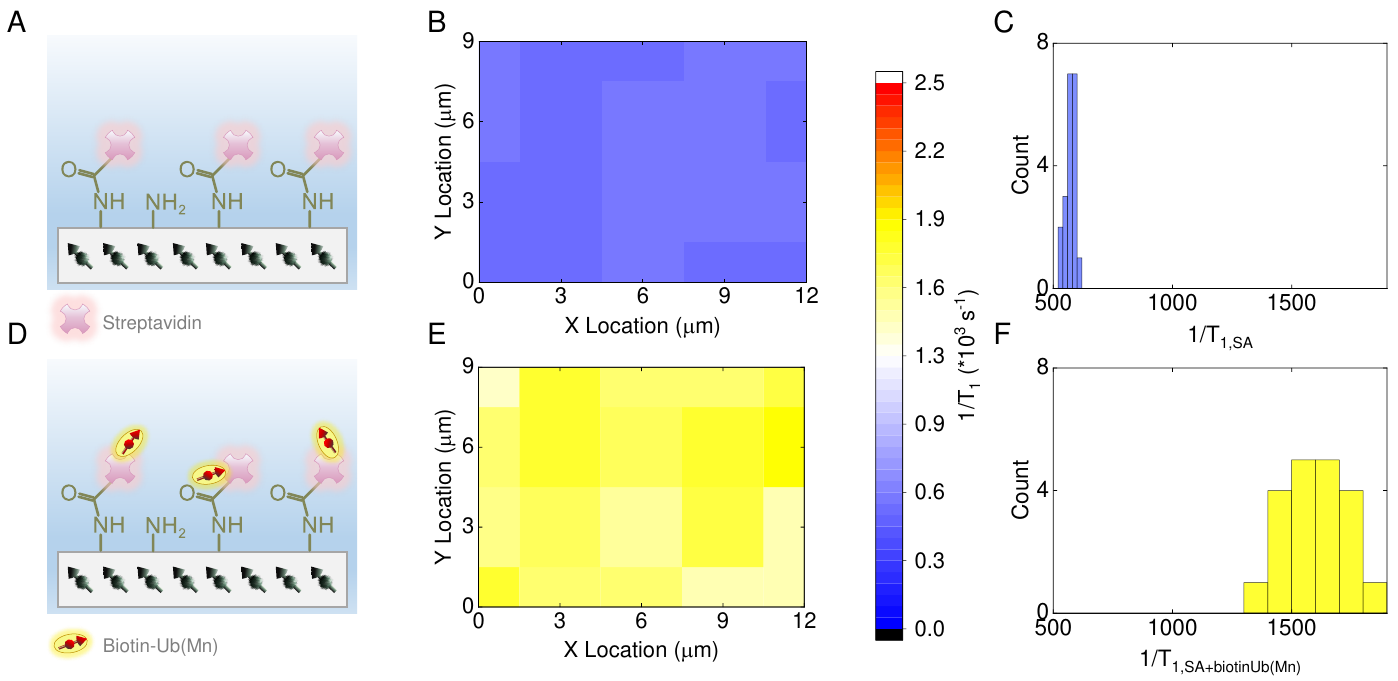}
\caption{\label{figs:Interaction in solution} {\textbf{Molecular interaction detection under solution conditions.} The PEI functionalization process is the same as the main text. (A) Schematic of the measurement condition when the diamond was coated with SA and immersed in MES buffer. (B) $T_1$ rate scanning with only SA. (C) Histogram of relaxation rates in (B). The mean rate is $(0.57\pm0.02)\times10^3\ \si{s}^{-1}$. (D) Schematic of the measurement condition after reacting with biotin-Ub(Mn). (E) and (F) are relaxation measurement results. The mean rate is $(1.6\pm0.1)\times10^3\ \si{s}^{-1}$. The lower signal intensity compared to dry conditions in the main text results from a thicker PEI layer in solution and a different diamond sample used in this measurement.}
}
\end{figure}

We measured the interaction between SA and biotin-Ub(Mn) in the hydrated environment. The background relaxation rates $1/T_{1,\text{SA}}$ with an averaged value of $(0.57\pm0.02)\times10^3\ \si{s}^{-1}$ were scanned when the diamond was coated with PEI+SA and immersed in MES buffer (\autoref{figs:Interaction in solution}A-C). Then the biotin-Ub(Mn) solution was added and reacted for 40 minutes. After washing off the excess protein, the diamond was maintained as well in MES buffer. The evident mean relaxation acceleration of $1.0\times10^3\ \si{s}^{-1}$ indicates that our modification and detection methods remain effective under solution conditions.

It should be noted that the composition of the buffer, such as salt components and pH value, will affect the NVs' relaxation process. Basic condition (pH$>$7) can decrease $T_1$ of carboxylated nanodiamond from 1.2 ms to 0.6 ms due to the negative ion layer\cite{pH2019ACSnano}. Diamagnetic electrolytes such as NaCl (MgSO$_4$, AlCl$_3$, KNO$_3$) solution increase $T_1$ by a factor of 1.5 compared to water\cite{acsnano2023electrolytesT1}. Therefore, it is indeed necessary to consider the influence of buffer components on $T_1$ in kinetic studies. To decouple the impact of buffer on the relaxation signal caused by molecular interactions, the buffer condition should be kept consistent during the measurement. This is also a common requirement for other similar molecular interaction analysis methods (such as SPR). \clearpage

\subsection{PEI stability after long-term immersion in solutions}
\begin{figure*}[ht]
    \centering
\includegraphics[width=.65\textwidth]{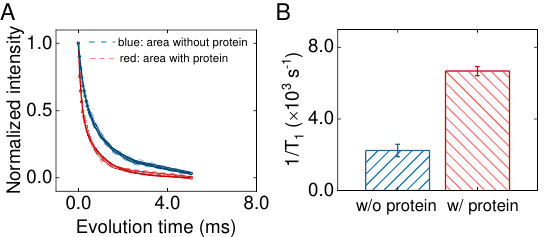}
\caption{\label{figs:PEIstability24h} {\textbf{PEI nanolayer stability test after 24-hour immersion in buffer.} The diamond was functionalized with a PEI nanolayer and then bonded with SA+biotin-Ub(Mn). After 24 h of immersion in MES buffer, the diamond surface was dried and proteins in a certain region were removed using AFM. (A) $T_1$ curves measured in regions with (4 lines) / without (5 lines) proteins. The dashed lines are experimental curves and the hollow dots are averaged data. The solid lines are fitting curves to the biexponential function. (B) Comparison of the averaged relaxation rate derived from the dashed lines in (A). The relaxation rate difference between the two regions is $(4.4\pm0.3)\times10^3\ \si{s}^{-1}$.} 
}
\end{figure*}

We tested the stability of PEI functionalization after long-term immersion in buffer. Specifically, the diamond sensor functionalized with PEI, SA, and biotin-Ub(Mn) was immersed in MES buffer for approximately 24 hours (\autoref{figs:PEIstability24h}). After drying, the measured signal intensity was $(4.4\pm0.3)\times10^3\ \si{s}^{-1}$, which is only 8$\%$ lower than the value reported in the main text. This little decrease confirms the robustness of this modification strategy over typical measurement timescales of several hours.

\clearpage

\bibliographystyle{unsrt}
\bibliography{Ref}